\documentclass[10pt,twocolumn,preprintnumbers,amsmath,amssymb,nofootinbib
,superscriptaddress]{revtex4-1}
\usepackage{graphicx,longtable,mathrsfs,color,array}
\usepackage[hidelinks]{hyperref}
\usepackage[usenames,dvipsnames]{xcolor} 
\usepackage{amssymb,amsmath,mathtools,mathrsfs,slashed} 
\usepackage{epsfig,subfigure,placeins,float} 
\usepackage{booktabs,longtable,ctable,multirow} 
\usepackage{exscale,relsize} 
\usepackage[normalem]{ulem} 
\usepackage{enumerate}
\usepackage{times, mathptmx} 
\usepackage{comment}
\usepackage{color}
\allowdisplaybreaks[1]

\begin{document}
\title{Cosmological Instabilities and the Role of Matter Interactions in Dynamical Dark Energy Models} 
\author{William J. Wolf}
\email{wwolf@uchicago.edu}
\affiliation{Physical Sciences Division, The University of Chicago, Chicago, IL 60637, USA}
\affiliation{Kavli Institute for Cosmological Physics, The University of Chicago, Chicago, IL 60637, USA}
\author{Macarena Lagos}
\email{mlagos@kicp.uchicago.edu}
\affiliation{Kavli Institute for Cosmological Physics, The University of Chicago, Chicago, IL 60637, USA}
\date{Received \today; published -- 00, 0000}

\begin{abstract}
We consider cosmological models with a dynamical dark energy field, and study the presence of three types of commonly found instabilities, namely ghost (when fields have negative kinetic energy), gradient (negative momentum squared) and tachyon (negative mass squared). In particular, we study the linear scalar perturbations of theories with two interacting scalar fields as a proxy for a dark energy and matter fields, and explicitly show how canonical transformations relate these three types of instabilities with each other. We generically show that low-energy ghosts are equivalent to tachyonic instabilities, and that high-energy ghosts are equivalent to gradient instabilities. 
Via examples we make evident the fact that whenever one of these fields exhibits an instability then the entire physical system becomes unstable, with an unbounded Hamiltonian. Finally, we discuss the role of interactions between the two fields, and show that whereas most of the time interactions will not determine whether an instability is present or not, they may affect the timescale of the instability. We also find exceptional cases in which the two fields are ghosts and hence the physical system is seemingly unstable, but the presence of interactions actually lead to stable solutions. These results are very important for assessing the viability of dark energy models that may exhibit ghost, gradient or tachyonic modes.
\end{abstract}

\date{\today}
\maketitle


\section{Introduction}\label{sec:introduction}

Our understanding of the universe is based on the $\Lambda$CDM model which agrees remarkably well with observational data \cite{Aghanim:2018eyx}, yet relies on the presence of major unknown components---dark matter and dark energy---which in turn play a crucial role in the evolution of the universe, making up currently $95\%$ of its total energy density. In particular, dark energy is assumed to be give by a cosmological constant, whose estimated observational value differs by more than 50 orders of magnitude from theoretical predictions \cite{Martin:2012bt}. For this reason, a number of alternative cosmological models have been proposed (see e.g.~\cite{Clifton:2011jh, Joyce:2014kja, Joyce:2016vqv, Ferreira:2019xrr} for  reviews), most of them promoting the cosmological constant to a dynamical field. However, the analysis of these models has shown that they are plagued by instabilities, which render them observationally unviable or fine tuned. 

In this paper, we consider general cosmological theories that propagate a dynamical dark energy field, and analyze the stability of linear cosmological perturbations. According to the standard scalar-vector-tensor decomposition of linear perturbations on a homogeneous and isotropic cosmological background \cite{Kodama:1985bj}, we focus on scalar perturbations only due to the fact that they are the most relevant observationally as they couple the matter density of the universe, and hence determine the fate of observables such as Cosmic Microwave Background temperature anisotropies and the evolution of large scale structures. In particular, we consider cosmological theories that propagate one dark energy scalar degree of freedom (DoF), and therefore include scalar-tensor (such as Horndeski \cite{Horndeski:1974wa, Deffayet:2009wt}, or DHOST \cite{Langlois:2015cwa, BenAchour:2016fzp}), vector-tensor (such as Generalized Proca \cite{Heisenberg:2014rta} or Generalized Einstein-Aether \cite{Zlosnik:2006zu}) and tensor-tensor models (such as dRGT massive gravity and bigravity \cite{deRham:2010kj,Hassan:2011zd}). 

We discuss the consequence of instabilities and show how typical instabilities discussed in the literature---namely ghost (associated to wrong signs of the kinetic energy of fields), gradient (wrong sign of squared momentum) and tachyon instabilities (wrong sign of squared mass)--- may exhibit themselves in different mathematical ways yet they are physically equivalent. The presence of any of these instabilities typically lead to a growth of cosmological perturbations, and hence understanding the properties of these instabilities is crucial for determining whether a cosmological model will be viable or not. At high energies, we would generically require models to be completely free of instabilities (although potential ways of exorcising high energy instabilities in Effective Field Theories have been discussed in \cite{ArkaniHamed:2003uy, Izumi:2007pb, Garriga:2012pk, Sbisa:2014pzo, Konnig:2016idp}). Nevertheless, at low energies, cosmological theories are allowed and need to have unstable perturbations in order to allow for the growth and formation of large-scale structure in the universe, although conditions on the growth timescales must be imposed in order to remain phenomenologically viable. Indeed, the analysis of stability and the conditions to reduce the viable parameter space of specific cosmological models has been extensively considered in the literature
(see e.g.~\cite{Cheung:2007st, Creminelli:2008wc, DeFelice:2011bh, Gubitosi:2012hu, Bloomfield:2012ff, Gleyzes:2013ooa, Piazza:2013pua, Kase:2014cwa, Konnig:2014xva, Gleyzes:2015pma, Konnig:2015lfa, DeFelice:2016uil,  DeFelice:2016ucp, DAmico:2016ntq, Frusciante:2016xoj, DeFelice:2017wel, Lagos:2017hdr, Heisenberg:2018mxx, Frusciante:2018vht, Crisostomi:2018bsp, Frusciante:2019xia}), where the positivity of all the coefficients of the action is imposed in the high energy limit, whereas at low energies conditions of the form $m\lesssim H_0$ on the mass $m$ of fields relative to the Hubble rate today $H_0$ are imposed. These stability conditions can also be used to efficiently search for the viable parameter space in numerical Einstein-Boltzman solvers \cite{EFTCAMB1, Zumalacarregui:2016pph}. 

In order to show explicitly the physical relation between the three types of instabilities, we consider canonical transformations which bring the physical system to a new mathematical basis while leaving physical properties invariant. We generically show that for systems with two coupled fields (as proxy for the dark energy field interacting with a matter field), low-energy ghosts are equivalent to tachyon instabilities, meanwhile high-energy ghosts are equivalent to gradient instabilities. The relationship between low-energy ghosts and tachyons was explored in \cite{Gumrukcuoglu:2016jbh} for the case of a massless canonical scalar field minimally coupled to general relativity, and here we generalize this to include models described by two interacting fields including both minimal and non-minimal couplings to gravity. Thus, for systems that describe scalar perturbations in many general cosmological models, these relationships exist and indicate that these instabilities correspond the same physical underlying entity regardless of which term or field they appear in. 

In addition, we discuss the timescale of instabilities and the role of field interactions. We find that most of the time, the presence of interactions will not determine whether an instability is present or not, although it does affect the timescale of the instability. This means that interactions must be carefully taken into consideration when determining whether a low-energy instability is phenomenologically viable or not. Additionally, we find that in certain cases interactions may lead to stable solutions in seemingly unstable system that do have negative energy contributions. Therefore, considering interactions carefully may help relax the stability conditions usually imposed at high energy in cosmological models, where positivity of all the coefficients in the action is required.

This paper is structured as follows. In Section \ref{sec:instabilities} we define the three types of instabilities typically referred to in the literature. We discuss how they appear in cosmology and their relevance. In Section \ref{sec:transformations} we summarize the concept of canonical transformations, and apply them to a toy model with a single scalar field to illustrate how this procedure evidences explicitly the relationship between the three instabilities. In Section \ref{sec:twofields} we consider realistic cosmological models with two interacting scalar perturbations, and generically show the effect of a canonical transformation. We discuss a number of simple cases of interest that illustrate the relation between the instabilities of two fields, as well as the role of interactions. Finally, in Section \ref{sec:discussion} we summarize our results and discuss their implications. Throughout this paper, we will be using natural units $c=M_P=1$, and metric signature $(-,+,+,+)$. 

\section{Classifying Instabilities}\label{sec:instabilities}
In this paper we will study the stability of scalar cosmological linear perturbations in models where there is one dark energy DoF. In general, the relevant quantity to consider will therefore be the quadratic action of linear perturbations around a time-evolving homogeneous and isotropic background metric:
\begin{equation}
ds^2=-dt^2+a(t)^2\eta_{ij}dx^idx^j,
\end{equation}
where $t$ is the physical time, $a(t)$ the scale factor that evolves in time, and $\eta_{ij}$ is the Minkowski spatial metric given by a 3D identity matrix. For gravity theories that propagate one dark energy field, we will generically have an action for two interacting real scalar perturbations: the dark energy perturbation and the matter perturbation (which would describe for instance the dark matter energy density). However, for simplicity, in this section we will consider an action for a single scalar cosmological perturbation and classify the different types of instabilities that this field can have. In the next section, we will extend our results to the realistic scenario of two interacting scalar perturbations. 

Let us consider for now a single DoF propagating on this cosmological background. In that case, for a real field $\phi$ with second-order derivative equations of motion, we can always write the action as:
\begin{equation}\label{LSingleCosmo}
S=\int d^4x\, \left[\frac{1}{2}  K_t(t)\dot{\phi}^2-\frac{1}{2}K_s(t) \eta^{ij}\partial_i\phi \partial_j\phi -\frac{1}{2}M(t)^2\phi^2\right],
\end{equation}
where overdots denote derivatives with respect to physical time, and $K_t$, $K_s$ and $M$ are three different coefficients that generically depend on time as they depend on the evolution of the cosmological background. Note that because the background breaks the explicit Lorentz space-time symmetry then we generically have that $K_t\not= K_s$. In addition, because the background does keep invariance under spatial rotations, then all the three spatial derivative components have the same coefficient $K_s$. 

From now on, we change the spatial coordinates to Fourier space according to:
\begin{equation}
\phi(t,\vec{x})=\frac{1}{(2\pi)^{3/2}}\int d^3k \, \phi_k(t,\vec{k})e^{i \vec{x}\cdot \vec{k}},
\end{equation}
where $\vec{k}$ is the vector wavenumber, and $k^2\equiv \eta^{ij}k_ik_j$. In Fourier space, we can calculate the Hamiltonian associated to the action in eq.~(\ref{LSingleCosmo}):
\begin{align}
H&= \int d^4x\,\left[ P\dot{Q} - \mathcal{L}  \right] \nonumber\\
&= \int d^3kdt\, \left[ \frac{1}{2}  K_t^{-1}P^2 + \frac{1}{2}K_s k^2Q^2 + \frac{1}{2}M^2Q^2   \right],\label{HSingleCosmo}
\end{align}
where we have defined the Hamiltonian variables as: $Q=\phi$ and $P=\partial \mathcal{L}/\partial \dot{\phi}$, with $\mathcal{L}$ the Lagrangian density of eq.~(\ref{LSingleCosmo}). Here we see that this Hamiltonian is expected to be positive definite when all $K_t$, $K_s$ and $M$ are positive, in which case the energy is bounded from below and hence the system is stable. 

In order to analyze explicitly the stability of the single-field system, let us consider a toy model where $K_t=\epsilon_t$, $K_s=\epsilon_s$, and $M^2=m^2\epsilon_m$, with all $\epsilon_A$ constants that can be positive or negative. The equation of motion for $\phi$ in this case is simply given by:
\begin{equation}\label{EqMotionToy}
\epsilon_t\ddot{\phi}_k+\left(\epsilon_s k^2 + \epsilon_m m^2\right)\phi_k=0.
\end{equation}

We say we have a stable field if all the coefficients $\epsilon_A=+1$, as in this case the solution to eq.~(\ref{EqMotionToy}) is given by $\phi_k(t) \propto \exp\{\pm it\sqrt{k^2+m^2}\}$, and hence the field just oscillates in time with a constant amplitude. In any other case, we will have an unstable field, with a Hamiltonian unbounded from below. We will distinguish three types of instabilities: we say we have a ghost if $\epsilon_g=-1$, a gradient instability if $\epsilon_k=-1$, and a tachyon if $\epsilon_m=-1$. In what follows, we discuss in detail these three different types of instabilities.

\subsection{Tachyon}
A tachyonic instability appears when the field has an effective negative mass squared, i.e.~when $\epsilon_m<0$. In particular, if $\epsilon_m=-1$ then the solution to eq.~(\ref{EqMotionToy}) is given by
\begin{equation}
\phi_k(t) \propto  e^{\pm it\sqrt{k^2-m^2}}.
\end{equation}
On the one hand, we see that when a tachyon is present then low-energy modes with $k\ll m$ are unstable and grow exponentially fast as $\phi_k\sim 
\exp\{mt\}$ in a fixed characteristic timescale  $t_{*}\sim m^{-1}$. Such a low energy instability is not a pathology in theories. Indeed, for cosmology this is found in General Relativity, and it is known as the Jean's instability, which is a necessary feature in order to form large scales structures in the universe \cite{1982ApJ...263L...1P}.
Nevertheless, this instability must be kept under control to avoid runaway amplitudes that would not be compatible with observations. In cosmology, we would require these low energy modes to grow on cosmological times scales, that is, $m\lesssim H_0$. 

On the other hand, we also see that high-energy modes with $k \gg m$ are insensitive to the instability and they simply oscillate as $\phi_k\sim \exp\{itk\}$. Therefore, we can have a perfectly well-defined effective field theory for scales $m<k<\Lambda$, with $\Lambda$ being the high-energy cut off scale of the theory. This is the case of General Relativity, where $\Lambda=M_P$. In this case, we have a cosmological model with a tachyonic instability, which is used to describe perturbation modes with $k$ such that $H_0< k < M_P$. 

\subsection{Gradient}
A gradient instability appears when the field has a negative momentum squared, i.e.~when $\epsilon_s<0$. In particular, if $\epsilon_k=-1$ then the solution to eq.~(\ref{EqMotionToy}) is given by
\begin{equation}
\phi_k(t)\propto e^{\pm it\sqrt{-k^2+m^2}}.
\end{equation}
In this case we see that low-energy modes with $k\ll m$ are stable and they simply oscillate with the same frequency as $\phi_k\sim \exp\{itm\}$. On the other hand, high-energy modes with $k\gg m$ grow exponentially fast as $\phi_k\sim \exp\{kt\}$ with a $k$-dependent characteristic timescale of $t_{k*}\sim k^{-1}$.
In this case, the modes that grow faster are those with higher $k$. If the theory in consideration is valid up to arbitrarily large energy scales, then we would have a high-energy instability. Such instabilities have been studied before, where it has been found that a quantum vacuum state would be unstable and would decay into particles exponentially fast \cite{Sbisa:2014pzo}. 

For an effective field theory with a cutoff $\Lambda$, the model could be pathological. Indeed, we would be interested in modes with $k\ll \Lambda$ which grow on timescales $t_{k*}\gg t_\Lambda$, and we would find that in the timescales of interest the modes close to the cutoff would be the largest ones and hence the effective  field theory may be ill-defined \cite{Trodden:2016zcu} (unless the high-energy completion of the EFT breaks Lorentz-breaking or locality-breaking).

\subsection{Ghost}
A ghost appears when the field has negative kinetic energy, i.e.~when $\epsilon_t<0$. In particular, if $\epsilon_t=-1$ then the solution to eq.~(\ref{EqMotionToy}) is given by
\begin{equation}
\phi_k(t)\propto e^{\pm t\sqrt{k^2+m^2}}.
\end{equation}
Here we see that both low-energy and high-energy modes are unstable and grow exponentially fast, similarly to the two previously discussed cases. For the same reasons mentioned before for gradient instabilities, a model with a high-energy instability like a ghost may not be well-defined even in the context of an effective field theory. 

We clarify that in the literature ghosts are commonly defined as fields with $\epsilon_t=\epsilon_s<0$ (e.g.~\cite{Sbisa:2014pzo}) as most models considered are Lorentz invariant. However, because we are interested in cosmological models where Lorentz invariance is broken, in this paper we define ghosts are fields with negative kinetic energy associated to time derivatives only. In this context, we say that a field with both $\epsilon_t$ and $\epsilon_s$ negative will have a ghostly gradient instability. In such a case, the equation of motion will be equivalent to that of a tachyon instability (modulo an overall minus sign), and thus the solution will not exhibit any high-energy instability, and it will be the same as that for a tachyonic field. This is not problematic as long as there are no couplings with additional fields, as we will see in the next sections. 

\subsection{Relations}
In this section we discuss the physical relationship between the three previously defined instabilities. From the action in eq.~(\ref{LSingleCosmo}) we see that the three type of instabilities are mathematically different: ghosts come from negative time derivatives, gradient instabilities come from negative spatial derivatives, and tachyons from negative non-derivative interactions.

Nevertheless, from the three solutions for $\phi_k(t)$ previously discussed we can see that a model with a physical instability for high-energy modes could have either a ghost or a gradient instability and in both situations the field would have the same solution $\phi_k(t)\sim \exp\{kt\}$. Similarly, a model with a physical low-energy instability could have either a ghost or a tachyon instability, and in both situations the field would have the same solution $\phi_k(t)\sim \exp\{ mt\}$. 

These relations can straightforwardly be understood by looking at the equation of motion (\ref{EqMotionToy}) and noticing that we can always multiply the equation by a factor $(-1)$ and the dynamics will not be affected. 
Explicitly, at high energies ($k\gg m$) the mass term becomes negligible and the equation of motion becomes $\ddot{\phi}_k-k^2\phi_k=0$ whether we have a ghost or a gradient instability. Similarly, at low energies ($k\ll m$) the equation of motion becomes $\ddot{\phi}_k-m^2\phi_k=0$ whether we have a ghost or a tachyonic instability.

We also mention the case of multiple instabilities. If we had for instance $\epsilon_t=\epsilon_s=\epsilon_m=-1$, then the equation of motion can be multiplied by an overall minus sign, and solutions would be stable. In this case, the Hamiltonian would be negative definite (and thus bounded from above), and since energy is conserved, solutions will be stable.

On the other hand, the fact that some instabilities are equivalent cannot be so easily understood at the level of the action. For instance, a rescaling of the field 
$\phi$ in the action (\ref{LSingleCosmo}) cannot change the sign of the individual three terms. For this reason, in the next section we show explicitly how to understand the relationship between the three instabilities at the level of the action. We will show that this can be achieved by performing canonical transformations to the associated Hamiltonian in eq.~(\ref{HSingleCosmo}).

Before studying canonical transformations, we emphasize that the results presented here are valid for theories more general than that of eq.~(\ref{LSingleCosmo}). Indeed, we will consider a generic scalar field with any number of spatial derivatives (including non-local terms). In Fourier space, we allow actions of the form
\begin{equation}\label{LSingleCosmok}
S=\int d^3k dt\, \left[\frac{1}{2}  K(t,k)\dot{\phi}_k^2  -\frac{1}{2}M(t,k)^2\phi^2_k\right],
\end{equation}
where now the functions $K$ and $M$ can have an arbitrary dependence on the wavenumber $k$ as well as time. Note that we did not write explicitly a term $K_s(t,k)$ as in Fourier space it can be reabsorbed into the definition of $M(t,k)$. We will say we have a high-energy (low-energy) ghost when $K<0$ in the limit of $k\rightarrow \infty$ ($k\rightarrow 0$), and a gradient (tachyon) instability when $M^2<0$ in the limit of $k\rightarrow \infty$ ($k\rightarrow 0$).

In the remaining of this paper, we focus on studying high and low energy instabilities which affect the early and late time cosmological evolution, leaving aside discussions on stability during intermediate scales as whether they are allowed or not must be analyzed case by case depending on how long they last and how fast perturbations grow when they are present.

\section{Single Field Instabilities}\label{sec:transformations}
In this section we briefly review canonical transformations, apply them to the single scalar field action (\ref{LSingleCosmok}) and show the previously discussed relationship between the three types of instabilities. 

\subsection{Canonical Transformations}
A canonical transformation is a transformation of the Hamiltonian variables from $\{P, Q\}$ to $\{P',Q'\}$ such that physics is invariant. In particular, we will generically start with a Hamiltonian density $\mathcal{H}(P,Q)$ (and associated Lagrangian $\mathcal{L}(\phi)$), apply the transformation and obtain a new Hamiltonian $\mathcal{H}'(P',Q')$ whose associated Lagrangian $\mathcal{L}'(\phi')$ will only differ to the original Lagrangian $\mathcal{L}(\phi)$ by a total derivative term. Therefore, both $\mathcal{L}$ and $\mathcal{L}'$ will lead to the same Euler-Lagrange equations and describe the same physical system. As a consequence, canonical transformations also preserve the Hamiltonian equations: 
\begin{align}
&\dot{P}=-\frac{\partial \mathcal{H}}{\partial Q}, \quad \dot{Q}=\frac{\partial \mathcal{H}}{\partial P},\nonumber\\
& \dot{P}'=-\frac{\partial \mathcal{H}'}{\partial Q'}, \quad \dot{Q}'=\frac{\partial \mathcal{H}'}{\partial P'},\nonumber
\end{align}
as well as the Poisson brackets:
\begin{align}
& \{ P,P\}= \{Q, Q\}=0, \quad \{ Q, P \}=1,\nonumber\\
& \{ P',P'\}= \{Q', Q'\}=0, \quad \{ Q', P' \}=1. \label{PoissonPrime}
\end{align}
The simplest canonical transformations one can consider are linear in the variables:
\begin{equation}\label{CanTransLin}
Q'= a_1Q+ a_2P, \quad P'= b_1Q + b_2P, 
\end{equation}
where $a_i$ and $b_i$ are arbitrary functions of time and $k$, such that eq.~(\ref{PoissonPrime}) is satisfied 
\begin{equation}
\{Q',P'\}= a_1b_2-a_2b_1=1.
\end{equation}
We therefore see that the most general linear canonical transformation for a single field will have three arbitrary independent functions of $t$ and $k$.

In the reminder of this paper, we will only consider linear canonical transformations as they are simple but still carry enough freedom to allow us to show explicitly the relationship between the three instabilities. Additionally,  linear canonical transformations would be consistent with linear perturbation theory, and therefore can be directly applied to cosmological models.

We note that a canonical transformation can also be performed at the level of the action $S$ instead of $H$, although it cannot be done with a simple field redefinition (see e.g.~\cite{Gumrukcuoglu:2016jbh} for a specific example). 

\subsection{Application}
Let us start by considering the action in eq.~(\ref{LSingleCosmok}) and calculating its associated Hamiltonian. First, we define the phase space variables as:
\begin{equation}
Q=\phi_k, \quad P=\frac{\partial \mathcal{L}}{\partial \dot{\phi}_k}= K\dot{\phi}_k,
\end{equation}
and obtain the Hamiltonian density $\mathcal{H}$:
\begin{equation}
\mathcal{H}= \frac{1}{2}\frac{P^2}{K} + \frac{1}{2}M^2 Q^2.
\end{equation}
Next, we proceed to apply the linear canonical transformation of eq.~(\ref{CanTransLin}) to this Hamiltonian. 

Under this transformation the new Hamiltonian density becomes:
\begin{align}
\mathcal{H}'&= \frac{1}{2}\frac{\left(M^2Ka_2^2+a_1^2\right)}{K}P^{'2}-\frac{\left(M^2Kb_2a_2+a_1b_1\right)}{K}P'Q'   \nonumber\\
&+\frac{1}{2}\frac{\left(M^2Kb_2^2+b_1^2\right)}{K}Q^{'2} ,
\end{align}
whose associated Lagrangian density can be found by defining a new field $\phi'_k\equiv Q'$ and using the Hamiltonian equation of motion $\dot{Q}=\partial \mathcal{H}'/\partial P'$ to obtain $P'$ as a function of $\phi'_k$ and $\dot{\phi}'_k$. Explicitly, we obtain:
\begin{align}\label{LCanTransf}
\mathcal{L}'&= P'\dot{Q}'-\mathcal{H} \nonumber\\
=&\frac{1}{A}\left[\frac{1}{2}K\dot{\phi}^{'2}_k+\left(M^2Ka_2b_2+a_1b_1\right)\phi_k' \dot{\phi}'_k-\frac{1}{2}M^2\phi^{'2}_k\right] ,
\end{align}
where $A\equiv a_1^2+M^2Ka_2^2$.
From this final Lagrangian we can identify new kinetic energy and mass terms, analogous to eq.~(\ref{LSingleCosmok}):
\begin{equation}\label{SingleFieldTransformed}
K'\equiv \frac{K}{A},\quad M^{'2} \equiv \frac{M^{2}}{A}+\frac{d}{dt}\left(\frac{M^2Ka_2b_2+a_1b_1}{A}\right),
\end{equation}
where we have made an integration by parts in the term $\phi'_k \dot{\phi}'_k$ so that the new Lagrangian is given by:
\begin{equation}
\mathcal{L'}=\frac{1}{2} K' \dot{\phi}^{'2}_k - \frac{1}{2}M^{'2} \phi^{'2}_k.
\end{equation}

Given this final result we can analyze the effect of the canonical transformation on the instabilities of the field $\phi_k$. For concreteness, let us consider the simple example studied in Section \ref{sec:instabilities}, where all coefficients of the action and the canonical transformation are constants, and the field has only second-order derivatives. We thus assume:
\begin{equation}
K=\epsilon_t, \quad M^2= \epsilon_sk^2+\epsilon_m m^2,
\end{equation}
where $\epsilon_A$ and $m$ are constants.

\subsubsection{Low Energy}
We see that if we start with a low-energy ghost field, for instance if $\epsilon_t=-1$ and $\epsilon_s=\epsilon_m=+1$, then the new field $\phi'$ will have the following kinetic and mass terms:
\begin{align}
K'=&\frac{-1}{(a_1^2-a_2^2m^2)-a_2^2k^2}, \\ M^{'2}=&\frac{m^2+k^2}{(a_1^2-a_2^2m^2)-a_2^2k^2}.
\end{align}
First of all, we notice that in this simple scenario, the only relevant free quantities are $a_1$ and $a_2$, whereas $b_1$ and $b_2$ are only relevant if there is a time dependence on the coefficients.
 In the low-energy regime, $k\rightarrow 0$, we coefficients of the new action become:
\begin{equation}
K'\rightarrow \frac{-1}{(a_1^2-a_2^2m^2)},\quad M^{'2}\rightarrow \frac{m^2}{(a_1^2-a_2^2m^2)}. \\ 
\end{equation}
Here we see that for an appropriate choice of the coefficients $a_1$ and $a_2$ such that $(a_1^2-a_2^2m^2)<0$ then the new field will not have a low-energy ghost instability but instead will have a tachyon instability. 
We note that, as expected, since canonical transformations do not change the physics of the system, the time-scale of the instability is just given by $m$, as the dependence in $K'_t$ and $M$ on $a_1$ and $a_2$ will not affect the equations of motion. 

\subsubsection{High Energy}
For the same model as before, with $\epsilon_t=-1$ and $\epsilon_s=\epsilon_m=+1$, $\phi$ starts with a high-energy ghost instability, but the new action will have a high-energy limit, $k\rightarrow \infty$, such that:
\begin{equation}
K'\rightarrow \frac{1}{a_2^2k^2}, \quad M^{'2}\rightarrow \frac{-1}{a_2^2},
\end{equation}
and thus we see the new field will have a gradient instability instead of a ghost instability, as long as the canonical transformation has $a_2\not=0$. Also, as expected,  the timescale of the instability is given by $k^{-1}$ as the factor of $a_2^2$ goes not affect the equations of motion in this case.

\vspace{0.2cm}
As a summary, at low energies, the original field was a ghost and the new one is a tachyon. At high energies, the original field was a ghost and the new one has a gradient instability. This demonstrates that there is an equivalence between low energy ghosts and tachyons and between high energy ghosts and gradients in this simple model, and that this can be seen at the level of the action by making a canonical transformation as well as through solutions to the equation of motion. 

We emphasize that in this example the canonical transformation did not remove the negative coefficient that signaled the presence of an instability, but rather moved it to another coefficient in the action. Thus, in general, if we want to avoid all instabilities (as we do at high energies), we would impose separate positivity conditions on both the kinetic and gradient terms. 
However, we note that this example can be generalized by allowing time-dependent coefficients in the canonical transformation. In that case, depending on how the coefficients evolve in time, it could be possible to completely remove a single negative terms (e.g.~if we had started with $K<0$ and $M^2>0$, and then chosen $A<0$ together with an appropriate time evolution such that $M^{'2}>0$ in eq.~(\ref{SingleFieldTransformed})), in which case the solution would be expected to be stable. Whether this is possible or not must be analyzed case by case, and for this reason the most conservative approach to avoid instabilities would be to simply impose positivity on both kinetic and gradient terms separately. 

\section{Two Interacting Fields Instabilities} \label{sec:twofields}

In this section we generalize the results from the single-field action to the case of two fields. This describes realistic cosmological models for scalar perturbations where one of the degrees of freedom is the dark energy field, and the other one is a matter field. Thus, we will describe models such as Quintessence, Horndeski, DHOST, Generalized Proca, and massive gravity. 
All these models will have a theory of scalar linear cosmological perturbations described by a quadratic action of the following form in Fourier space (see \cite{Lagos:2017hdr} for a general proof and the explicit action for some scalar-tensor, vector-tensor and bimetric models):
\begin{align}\label{twofields}
S &= \int d^3k dt\, \left[\frac{1}{2}  K_1(t,k)\dot{\phi}_{1k}^2  -\frac{1}{2}M_1(t,k)^2\phi_{1k}^2 \right.  \nonumber\\
&\left.+D(t,k)\left(\Dot{\phi_{1k}}\phi_{2k}-\Dot{\phi_{2k}}\phi_{1k}\right)+ \frac{1}{2}  K_2(t,k)\dot{\phi}_{2k}^2  -\frac{1}{2}M_2(t,k)^2\phi_{2k}^2 \right]
\end{align}
where, as in the previous section, the coefficients $K_{1,2}$, $D$ and $M_{1,2}$ can be functions of time and wavenumber $k$. Their explicit forms will depend on the specific theory under consideration. We will keep working on Fourier space but, for simplicity, we will omit the subscript $k$ in the fields. We note that we have not included mass or kinetic mixing terms because one can always perform a field redefinition to bring the system in the form of eq.~(\ref{twofields}).

As before, we will briefly look at the equations of motion for this system and then consider a linear canonical transformation to demonstrate the physical relationship between the three types of instabilities. 

For illustrative purposes, let us for now assume that the coefficients of the action (\ref{twofields}) are constants in time. In that case, the equations of motion are simply given by:
\begin{align}
K_1\ddot{\phi}_{1}+M_1^2\phi_{1}=-2D\dot{\phi}_{2},\quad K_2\ddot{\phi}_{2}+M_2^2\phi_{2}=2D\dot{\phi}_{1},
\end{align}
where we explicitly see that the presence of the coupling term $D$ leads to an exchange of energy between the two fields, which will ultimately cause any instability to affect both fields. The explicit solutions to these equations of motion are linear combinations of four exponential terms as: 
\begin{align}\label{twofieldsol}
&\phi_{1,2}\propto \exp\left\{\pm it  \left(2K_1K_2\right)^{-1/2} \cdot \right. \nonumber\\
& \left. \sqrt{X \pm \sqrt{8D^2(X-2D^2)+(K_2M_1^2-K_1M_2^2)^2}}\right\}
\end{align}
where we have defined $ X \equiv 4D^2+M_2^2K_1+M_1^2K_2$.
We mention that here the proportionality factors have been omitted, but they depend on $K_{1,2}$, $D$, and $M_{1,2}$ and are such that for $D=0$ (decoupled system) then each field has only two solutions. Similarly to the previous section, we illustrate the behaviour of these solutions in a simple setting where $K_{1,2}= \epsilon_{t1,2}$, $M_{1,2}^2=k^2\epsilon_{s1,2}+ m_{1,2}^2\epsilon_{m1,2}$, where all $\epsilon$ quantities are constants that can take the values $\pm 1$, and $m^2_{1,2}>0$. 
Upon inspection of the solutions, it is clear that if all $\epsilon$ are positive then the system is stable. In particular, the solutions can be rewritten as:
\begin{align}
&\phi_{1,2}\propto \exp\left\{\pm  it  \sqrt{Y^2\pm \sqrt{Y^4-Z^2}}\right\}
\end{align}
where $Y^2\equiv4D^2+2k^2+m_1^2+m_2^2$ and $Z^2\equiv4k^2(k^2+m_1^2+m_2^2)+4m_1^2m_2^2$ are both always positive. These four exponential terms will thus always lead an oscillatory behaviour with constant amplitudes. 

We emphasize that, in general, the sign of the interaction term $D$ never affects the stability of the system because it always shows up as squared in the solution of eq.~(\ref{twofieldsol}). Even though the sign is irrelevant for determining the presence of instabilities, its absolute value is still important for determining the scale of any instabilities that may be present, as we will see next.

If any one of the coefficients $\epsilon$ is negative, then an instability will appear. For instance, if $\epsilon_{1t}=-1$ with all the other terms positive, then the solutions in the low and high-energy regimes will be dominated by real exponential terms as:
\begin{align}\label{two_field_limit}
& \phi_{1,2} \propto \exp\left\{t\sqrt{-Z+\sqrt{Z^2+4m_1^2m_2^2}}\right\},\, \text{for} \; k \rightarrow 0,\nonumber\\ 
& \phi_{1,2} \propto \exp\left\{kt\sqrt{\frac{4D^2}{k^2}+\sqrt{4+\frac{16D^4}{k^4}}}\right\},\, \text{for} \; k \rightarrow \infty,
\end{align}
where $Z\equiv m_2^2-m_1^2-4D^2$. For concreteness, in these solutions we have assumed that $D$ scales with $k$ in such a way that is relevant in the given regime (i.e.~$D\propto k$ for $k\rightarrow \infty$, and $D\propto k^0$ for $k\rightarrow 0$), although solutions for any other scaling for $D$ can straightforwardly also be obtained from eq.~(\ref{twofieldsol}).

From the solutions in eq.~(\ref{two_field_limit}) we explicitly see that due to the presence of a ghost instability in the field $\phi_1$ then the entire system is unstable as both fields exhibit growing exponential solutions. This is due to the presence of the interaction term $D$. In the limit $D\rightarrow 0$, the two fields are no longer described by linear combinations of four solutions, but rather each field is reduced to two solutions (as in the single-field example of the previous section) where the instability then only resides in field with the initial ghost.

We also mention the case of multiple instabilities. If one of the fields has a ghost and gradient instabilities then the entire system is generically unstable, as opposed to the single-field case where the evolution of the field was found to be oscillatory and stable. This is due to the presence of interactions. However, these equations also do show that it is possible to have stable solutions in situations where there are multiple instabilities for some specific cases. For example, the solutions when we have two ghosts, i.e.~$\epsilon_{t1}=\epsilon_{t2}=-1$ and all other terms positive are given below. As before, the solutions are given in the low and high energy limits and $D$ has been assumed to scale so that it is relevant in each energy regime:
\begin{align}\label{two_ghost_limit}
& \phi_{1,2} \propto \exp\left\{\pm it\sqrt{W \pm \sqrt{W^2-4m_1^2m_2^2}}\right\},\, \text{for} \; k \rightarrow 0, \nonumber \\ 
& \phi_{1,2} \propto \exp\left\{\pm kt\sqrt{2-\frac{4D^2}{k^2}\left(1\pm \sqrt{1-\frac{k^2}{D^2}}\right)}\right\},\, \text{for} \; k \rightarrow \infty.
\end{align}
Here $W=4 D^2-m_1^2-m_2^2$. We see that for $W>0$ and $ (W^2-4m_1^2m_2^2)\geq 0$ the solutions in the low-energy regime will be stable and oscillating. In this example, a seemingly unstable model with two ghosts turned out to be stable due to the presence of the interaction $D$ (note that if $D=0$ then the solutions for a model with two ghosts would always be unstable).

Similarly, we see that there are also cases in which the high-energy solutions are stable. If $D^2<1$, the system will be unstable while $D^2>1$ will result in stable, oscillating solutions (more generally, we should compare the size of $D^2$ to the size of the coefficients in the gradient terms of the action). In the high-energy case, it was important that $D\propto k$ in order to get stable solutions. Indeed, in eq.~(\ref{two_field_limit}), if $D\propto k^0$, the solutions will be unstable. In this case, both fields will have growing solutions proportional to $e^{kt}$.

The cases given in eq.~(\ref{two_ghost_limit}) are exceptional as we find that most of the time the size of $|D|$ does not determine whether an instability is present or not. However, in eq.~(\ref{two_field_limit}) we do find that the timescale of the instability does depend on $D$, as long as it is relevant on the given energy regime. Therefore, if we do allow the presence of instabilities in any given model then analyzing the role of $D$ will be crucial for imposing conditions on the free parameters of the model that make the instability stay under control. For instance, for low-energy instabilities on cosmological models, we would impose a condition on the masses and $D$ such that the timescale of the instability is of the order of the Hubble rate or smaller.

Finally, the same general relations between types of instabilities found for the case of single field hold in this interacting system. In particular, we see that there will be an instability present in the high-energy regime if either a field has a high-energy ghost or a gradient instability. However, the timescale of the instability may be different in these two cases if the interaction term dominates the evolution. Similarly, in the low-energy regime, there will be an instability if a field has low-energy ghost or a tachyon (with an instability timescale that may change again). Thus, the relations we observed in the previous section are shown to generalize to this more realistic scenario of two interacting fields.

\subsection{Application}
Proceeding as in the previous section, we now define and carry out canonical transformations on eq.~(\ref{twofields}). First, we define our phase space variables and calculate the associated Hamiltonian. Then we will calculate the new Hamiltonian under the defined transformation. The canonical variables are:
\begin{align}
Q_1=\phi_1, \quad P_1=\frac{\partial \mathcal{L}}{\partial \dot{\phi}_1}= K_1\dot{\phi}_1 +D\phi_2, \\
Q_2=\phi_2, \quad P_2=\frac{\partial \mathcal{L}}{\partial \dot{\phi}_2}= K_2\dot{\phi}_2-D\phi_1,
\end{align}
with a Hamiltonian density given by:
\begin{align}\label{twofieldH}
\mathcal{H} &=\frac{1}{2}\frac{P_1^2}{K_1}+\frac{1}{2}\frac{P_2^2}{K_2}+\frac{1}{2}M_1^2Q_1^2+\frac{1}{2}M_2^2Q_2^2
\nonumber \\ &+\frac{1}{2}\frac{D^2 Q_2^2}{K_1}+\frac{1}{2}\frac{D^2 Q_1^2}{K_2}-\frac{D P_1 Q_2}{K_1}+\frac{D P_2 Q_1}{K_2}.
\end{align}
An interesting feature of this Hamiltonian is that if we write it in terms of Lagrangian variables, the terms that involve mixing between the field disappear. Explicitly, we find:
\begin{equation}
\mathcal{H}=\frac{1}{2} \left(\text{$K_1$} \text{$\Dot{\phi_1}$}^2+\text{$K_2$} \text{$\Dot{\phi_2}$}^2+\text{$M_1$}^2 \text{$\phi_1 $}^2+\text{$M_2$}^2 \text{$\phi_2 $}^2\right).
\end{equation}
The presence of $D$ vanishes due to the antisymetric structure of the mixing terms $\Dot{\phi_1}\phi_2$ and $\Dot{\phi_2}\phi_2$. This makes explicitly again the fact that the sign and size of $D$ does not affect the stability of the system most of the time, as it does not affect the sign of the Hamiltonian (and whether it is positive definite). The Hamiltonian will only depend on the sings of $K_{1,2}$ and $M_{1,2}$. For this reason, in the results that will follow we will always write mixing terms in an antisymmetric form and analyze the signs of the kinetic and mass terms only. Furthermore, as we shall see below, in order to determine whether or not there is an instability it is important for the kinetic and mass terms to be in a diagonal form as eq.~(\ref{twofields}), that is, with no kinetic or mass interactions of the form $\dot{\phi}_1\dot{\phi}_2$ nor $\phi_1\phi_2$.

Next we will carry out two transformations to this Hamiltonian to highlight different aspects of the relations between instabilities in this type of system. We will consider transforming both fields in a manner similar to what we did in the first section and show that this makes the analysis we did on the equations of motion explicit at the level of the action.
The transformation is given by:
\begin{align}\label{CanTransLin2}
&Q'_1= a_1Q_1 + a_2P_1, \quad P'_1= b_1Q_1 + b_2P_1, \nonumber\\
&Q'_2= c_1Q_2 + c_2P_2,  \quad P'_2= d_1Q_2 + d_2P_2.
\end{align}
While the following conditions as well as eq.~(\ref{PoissonPrime}) must be satisfied for a canonical transformation:
\begin{align}
\{Q'_1,P'_1\}= a_1b_2-a_2b_1=1, \nonumber\\
\{Q'_2,P'_2\}= c_1d_2-c_2d_1=1.
\end{align}
As before, we will carry out the transformation to obtain the new Hamiltonian and explore what relations this shows between the types of instabilities we have previously defined.

We then proceed from the original Hamiltonian eq.~(\ref{twofieldH}) and use our transformation eq.~(\ref{CanTransLin2}) to obtain the new Hamiltonian which has the following form:
\begin{align}
	\mathcal{H'} &= \Gamma_1 P_1'^2 + \Gamma_2 P'_1Q'_1 + \Gamma_3 Q_1'^2 + \Gamma_4 P_2'^2 + \Gamma_5 P'_2Q'_2 + \Gamma_6 Q_2'^2 
	\nonumber\\  &+  \Gamma_7 Q_1'P'_2 + \Gamma_8 P_1'Q'_2 +  \Gamma_9 Q'_1Q'_2 +  \Gamma_{10} P_1'P'_2,
\end{align}
where the $\Gamma$s are terms comprised of coefficients from the transformation and the original Hamiltonian. From here, we can explicitly see the need for the Hamiltonian (and hence Lagrangian) to be in a diagonal form in order to study its stability, as mixing terms cannot generically be assumed to be positive definite even if the coefficient in front is positive. For example, $P_2'^2$ will always be positive, but that still leaves open the possibility of introducing additional minus signs through a term multiplied by $P_1'P_2'$. Therefore, we will need to make our final Lagrangian diagonal before making determinations about the relationship between different types of possible instabilities.

As before, we then use Hamilton's equation $\dot{Q'_1}=\partial \mathcal{H}'/\partial P_1'$ and $\dot{Q'_2}=\partial \mathcal{H}'/\partial P_2'$ to obtain the new kinetic terms for the transformed fields $\phi_1'$ and $\phi_2'$ in terms of the Hamiltonian variables. We also make use of the Legendre transform to find the transformed Lagrangian and switch back to using Lagrangian variables:
\begin{align}\label{transfLtwofields}
 \mathcal{L'} &= \frac{C_1}{2A}\Dot{\phi_1'}^2 + \frac{C_2}{A}\Dot{\phi_1'}\phi_1' - \frac{C_3}{2A} \phi_1'^2 + \frac{C_4}{2A}\Dot{\phi_2'}^2 + \frac{C_5}{A}\Dot{\phi_2'}\phi_2' 
    \nonumber\\ & -\frac{C_6}{2A} \phi_2'^2 + \frac{C_7}{A}\Dot{\phi_1'}\Dot{\phi_2'} +  \frac{C_8}{A}\Dot{\phi_1'}\phi_2' +  \frac{C_9}{A}\Dot{\phi_2'}\phi_1' +  \frac{C_{10}}{A}\phi_1'\phi_2'.
\end{align}
We have defined A:
\begin{align}
A&\equiv a_2^2 \left(c_2^2 \left(D^2+k_2 m_1^2\right) \left(D^2+k_1 m_2^2\right)+c_1^2 k_1 m_1^2\right) \nonumber \\
&+2 a_1 a_2 c_1 c_2 D^2+a_1^2 \left(c_2^2 k_2 m_2^2+c_1^2\right).
\end{align}
Here the $C$s are functions of our transformation coefficients which we will assume are all constants in time and are given in Appendix \ref{App:twofieldcoeffs}. Now, we proceed to consider specific examples and study the stability of this new Hamiltonian. Before that, we note that this Hamiltonian can be simplified by making integration by parts on the terms $\Dot{\phi_1'}\phi_1'$ and $\Dot{\phi_2'}\phi_2'$ and reabsorbing them into their corresponding mass terms. However, if the coefficients of these terms are constants in time, they simply lead to boundary terms that can be ignored. Furthermore, we also bring into an antisymmetric form the terms $\Dot{\phi_1'}\phi_2'$ and $\Dot{\phi_2'}\phi_1'$ by performing integration by parts, as shown in Appendix \ref{App:AntisymmetricTerms}. This will allow us to ignore these terms most of the time in the stability analysis as previously discussed.

In what follows, we will choose the coefficients of the canonical transformations such that there are no kinetic and mass mixing terms (involving $\Dot{\phi_1'}\Dot{\phi_2'}$ and $\phi_1'\phi_2'$) in $\mathcal{L}'$, which will allow us to easily study the stability of the system by simply looking the the signs of the diagonal kinetic and mass terms. While we could do this using only the canonical transformation we have defined, we introduce an additional field redefinition at this stage because it will give us additional freedom to be fully general in illustrating the relationships between ghost, gradient and tachyon instabilities in this physically realistic system. 

The field redefinition is given below where $\phi_3'$ and $\phi_4'$ are linear combinations of the transformed fields, and $\alpha_{1,2}$ and $\beta_{1,2}$ are some a priori arbitrary coefficients. 
\begin{align}\label{FieldRedef}
&\phi_1' = \alpha_1 \phi_3' + \alpha_2 \phi_4', \quad \phi_2' = \beta_1 \phi_3' + \beta_2 \phi_4'.
\end{align}
After making the field redefinition, we will later choose $\alpha_1$ and $\alpha_2$ such that they diagonalize the kinetic and mass matrices thereby eliminating problematic mixing terms. First we will show that this reproduces the results of the single field example from the beginning by demonstrating that a low energy ghost can generically be exchanged for a tachyon and that a high energy ghost can be generically exchanged for a gradient. Then we will explore some further properties. 

We will carry out the analysis of this Lagrangian in detail using the following assumptions. For concreteness, similarly to previous discussions, we assume that the initial coefficients of the action have the following form: $K_i=\epsilon_{it}$ and $M_i^2= \epsilon_{is}k^2+\epsilon_{im} m_i^2$, where i represents either $\phi_1$ or $\phi_2$, all the $\epsilon$ quantities can take the values $\pm 1$ and $m_i^2$ are assumed to be constants. Furthermore, for illustrative purposes, we  assume $D$ to be constant in time and independent of $k$, although such an assumption can be easily generalized. Additionally, in the analysis that follows, any coefficients from the canonical transformation and field redefinition are assumed to be constants in time and independent of $k$ (except for $\alpha_{1,2}$ that are chosen to diagonalize mass and kinetic interactions, which leads to a $k$ dependence on them, while they indeed remain time independent).

\subsection{Example I: Ghosts to Tachyons and Gradients}
Using the freedom we have with the canonical transformation, we make the choice $a_1=c_2=0$ with our transformation coefficients in order to simplify the resulting Lagrangian and make the relationships between various instabilities more evident. The resulting Lagrangian can be written in the following way:
\begin{align}\label{example1L}
\mathcal{L'} &=\frac{x\left[x-z \right]}{16D^2c_1^2M_1^2} \dot{\phi}_3'^2-\frac{x\left[x+y \right]}{16D^2c_1^2K_1} \phi_3'^2 \nonumber\\ 
&+\frac{x\left[x+z \right]}{16D^2c_1^2M_1^2} \dot{\phi}_4'^2-\frac{x\left[x-y \right]}{16D^2c_1^2K_1} \phi_4'^2. 
\end{align}
Note that we have omitted terms that can be integrated out and lead to boundary terms when we have constant coefficients. Also, we have defined:
\begin{align}
&x=\left(K_2^2M_1^4+2K_2M_1^2(4D^2-K_1M_2^2)+(4D^2+K_1M_2^2)^2\right)^{1/2},\nonumber \\ 
&y=(4D^2+K_2M_1^2-K_1M_2^2), \quad z = (4D^2-K_2M_1^2+K_1M_2^2).
\end{align}
Here we see that this transformation is only valid
when the total terms inside the squared root of $x$ are positive, in order to keep the fields real.
Additionally, we note that with this canonical transformation, field redefinition and particular choice of coefficients there is no $D'$ term. Therefore, this system has been explicitly decoupled, and there are no interactions between $\phi_3'$ and $\phi_4'$ anymore.
However, the system is still physically equivalent to the old one as the equations of motion for $\phi_3'$ and $\phi_4'$ leads to equivalent solutions with same frequencies given in eq.~(\ref{twofieldsol}) that were obtained from analyzing the system with $\phi_1$ and $\phi_2$.

In order to illustrate explicitly what canonical transformations can do to a two-field action, let us now consider a scenario where we have an initial ghost in the field $\phi_1$, that is, $\epsilon_{1t}=-1$, $\epsilon_{2t}=+1$ and $\epsilon_{is}=\epsilon_{im}=+1$. Next, we take the low and high energy limits of the transformed Lagrangian, assuming that the free coefficients left in the canonical transformations are independent of $k$.

We will now analyze the stability of this system eq.~(\ref{example1L}) which is equivalent to the initial Lagrangian (\ref{twofields}), but with new coefficients $K'_{3,4}$, $M'_{3,4}{}^{2}$ for the fields $\phi_3'$ and $\phi_4'$. This stability analysis is done, as in the previous section, by looking at the sign of the kinetic and mass coefficients in the low and high-energy regimes.

\subsubsection{Low Energy} 
In the low-energy regime, $k \to 0$, we find the following kinetic and mass coefficients:
\begin{align}
&K_3'\rightarrow \frac{x\left[\sqrt{z^2+16D^2m_1^2}-z\right]}{16D^2c_1^2m_1^2}, \nonumber \\ 
&M_3^{'2}\rightarrow \frac{x\left[y-\sqrt{y^2-16D^2m_2^2}\right]}{16D^2c_1^2}, \nonumber \\
&K_4'\rightarrow \frac{x\left[\sqrt{z^2+16D^2m_1^2}+z\right]}{16D^2c_1^2m_1^2}, \nonumber \\
&M_4^{'2}\rightarrow \frac{-x\left[\sqrt{y^2-16D^2m_2^2}+y\right]}{16D^2c_1^2}.
\end{align}
Here $x=\left(m_1^4+2m_1^2(4D^2+m_2^2)+(4D^2-m_2^2)^2\right)^{1/2}$, and is always real. Also, we have defined  $y=(4D^2+m_1^2+m_2^2)$ and $z=(-4D^2+m_1^2+m_2^2)$, in order to easily determine if there is a minus sign in the kinetic and mass terms, and thus if there is an instability. We find that $M_4^{'2}<0$, and hence we transform a low-energy ghost into a tachyon while all of the other coefficients in this new, but physically equivalent system are manifestly positive. Furthermore, this results in unstable solutions corresponding to the low-energy limit of eq.~(\ref{two_field_limit}). 

Analogously, we mention that if we had started with a ghost in $K_2$ instead ($\epsilon_{2t}=-1$), the new action would have kept one ghost in $\phi_3'$ and would exhibit no tachyons. 

We emphasize that because $\phi_{3}'$ and $\phi_4'$ are linear combinations of the initial fields $\phi_1$ and $\phi_2$ and their derivatives, it is not possible to obtain a direct relationship between instabilities in the initial and the final fields. Indeed, an initial ghost in $\phi_1$ is transformed to a tachyon in $\phi_4'$, whereas a tachyon in $\phi_1$ is transformed to a ghost in $\phi_3'$.
Thus, we can clearly see that the entire system is unstable, and instabilities are the same physical entity regardless of the explicit term they appear in within the Lagrangian, and regardless of the field in which they appear within an interacting system. 

This example generalizes the results found for a single field in the previous section. Here we show that we can also transform low-energy ghosts into tachyons, even in realistic scenarios when interactions are present.

\subsubsection{High Energy}
In the limit $k \to \infty$ we obtain the following kinetic and mass terms in the transformed action:
\begin{align}
& K_3'\rightarrow \frac{1}{c_1^2}, \quad M_3^{'2}\rightarrow \frac{k^2}{c_1^2}, \nonumber \\ 
& K_4'\rightarrow \frac{k^2}{D^2 c_1^2}, \quad M_4^{'2}\rightarrow  -\frac{k^4}{D^2 c_1^2}. 
\end{align}
Here again the canonical transformation has allowed us to move a minus sign in the kinetic term of the action into the mass term of the action. 
In particular, we see that the initial high-energy ghost in $\phi_1$ now appears as a gradient in $\phi_4$. Because we have assumed that $D\propto k^0$, then $D$ does not affect the  timescale of the instability and the solutions go as $e^{kt}$ for the field that contains the instability.

We again conclude that the results found for the single field generalize to this more realistic scenario. 

\subsection{Example II: Two Ghosts} 
Here we consider an example where the action started initially with two ghosts. In this case, we do not perform the field redefinition in eq.~(\ref{FieldRedef}), and instead use the freedom only from the canonical transformation to diagonalize mass and kinetic interactions. In particular, we choose $a_1=c_1=0$,
which gives the following Lagrangian:
\begin{align}\label{example3L}
\mathcal{L'} &=+\frac{K_2 \dot{\phi }_1'^2}{2 a_2^2 \left(D^2+K_2 M_1^2\right)}-\frac{M_2^2 \phi _1'^2}{2 a_2^2 \left(D^2+K_1M_2^2\right)}\nonumber \\ 
&+\frac{K_1 \dot{\phi }_2'^2}{2 c_2^2 \left(D^2+K_1 M_2^2\right)}-\frac{M_1^2 \phi _2'^2}{2 c_2^2 \left(D^2+K_2 M_1^2\right)} \nonumber\\ 
&+D'(\dot{\phi_1'}\phi_2-\dot{\phi_2}\phi_1),
\end{align}
where $D'$ is given by:
\begin{align}
D'=\frac{D(2D^2+K_1M_2^2+K_2M_1^2)}{2a_2c_2(D^2+K_1M_2^2)(D^2+K_2M_1^2)}.
\end{align}
Next, we analyze the stability of the action when $\epsilon_{1t}=\epsilon_{2t}=-1$ and $\epsilon_{is}=\epsilon_{im}=+1$ and look separately at the low and high-energy regimes.

\subsubsection{Low Energy} 
In the low-energy regime, $k \to 0$, we find the following kinetic and mass coefficients:
\begin{align}
& K_3'\rightarrow -\frac{1}{a_2^2 \left(D^2-m_1^2\right)}, \quad M_3^{'2}\rightarrow \frac{m_2^2}{a_2^2 \left(D^2-m_2^2\right)}, \nonumber \\ 
& K_4'\rightarrow -\frac{1}{c_2^2 \left(D^2-m_2^2\right)}, \quad M_4^{'2}\rightarrow  \frac{m_1^2}{c_2^2 \left(D^2-m_1^2\right)}. 
\end{align}
Here we see that depending on the relative size of $D^2$ and $m^2_{1,2}$, the negative signs will appear in different terms. For instance, if $D^2<m_{1,2}^2$, then the two initial low-energy ghosts appear now as two tachyons.
This will lead to unstable solutions as this choice will cause eq.~(\ref{two_ghost_limit}) to be dominated by real exponentials. 
If $D^2$ is larger than one of the masses but not the other then we would have a mix of ghosts and tachyons that would likewise correspond to unstable solutions in eq.~(\ref{two_ghost_limit}).

On the other hand, if $4D^2>(m_1^2+m_2^2)$ and $(D^2-m_1^2-m_2^2)^2>4m_1^2m_2^2$, which are the same conditions as $W>0$ and $W^2-4m_1^2m_2^2>0$ mentioned previously, there will still be two ghosts in the action; however, this will then correspond to the low energy limit of eq.~(\ref{two_ghost_limit}) with stable, oscillating solutions. 

\subsubsection{High Energy} 
In the limit $k \to \infty$ we obtain the following kinetic and mass terms in the transformed action:
\begin{align}
& K_3'\rightarrow \frac{1}{a_2^2 k^2}, \quad M_3^{'2}\rightarrow -\frac{1}{a_2^2},  \\ \nonumber
& K_4'\rightarrow \frac{1}{c_2^2 k^2}, \quad M_4^{'2}\rightarrow  -\frac{1}{c_2^2}. 
\end{align}
In this case, we see that the two initial ghosts appear now as two gradients. Similarly to the previous example, since we have assumed $D\propto k^0$, the two fields are effectively decoupled, and the solutions correspond simply to growing exponentials of the form $e^{kt}$.
 
\subsection{Example III: Stable Two-Ghost System}
Now, we analyze further the case of two ghosts with stable solutions, bringing the Lagrangian into a form where it is easily seen that the system is stable. We will make the same choice of coefficients as in Example I for the canonical transformation and field redefinition. 
We start by using eq.~(\ref{example1L}) and considering $\epsilon_{1t}=\epsilon_{2t}=-1$ and $\epsilon_{is}=\epsilon_{im}=+1$. 

\subsubsection{Low Energy} 
In the low energy limit, $k \to 0$, we find the following kinetic and mass coefficients:
\begin{align}
&K_3'\rightarrow \frac{x\left[\sqrt{z^2-16D^2m_1^2}+z\right]}{16D^2c_1^2m_1^2}, \nonumber \\ 
&M_3^{'2}\rightarrow \frac{x\left[y-\sqrt{y^2-16D^2m_2^2}\right]}{16D^2c_1^2}, \nonumber \\
&K_4'\rightarrow -\frac{x\left[z-\sqrt{z^2-16D^2m_1^2}\right]}{16D^2c_1^2m_1^2}, \nonumber \\ 
&M_4^{'2}\rightarrow -\frac{x\left[\sqrt{y^2-16D^2m_2^2}+y\right]}{16D^2c_1^2}.
\end{align}
Now $x=\left(-2 m_1^2 \left(4 D^2+m_2^2\right)+\left(m_2^2-4 D^2\right)^2+m_1^4\right)^{1/2}$, $y=(4D^2-m_1^2+m_2^2)$ and $z=(4D^2+m_1^2-m_2^2)$. Notice that x is now real for restricted values of the coefficients $D$, $m_1$ and $m_2$ and again must be so in order to have a valid transformation because the fields in our initial system are real. These restricted values correspond to $W^2-4m_1^2m_2^2>0$. Furthermore, if $y$ and $z$ are both positive, then all of these conditions become equivalent to those for which we could obtain stable solutions with two initial ghosts in eq.~(\ref{two_ghost_limit}).
Considering these values, we see that $K_3'$ and $M_3'^2$ are both positive while  $K_4'$ and $M_4'^2$ are both negative. However, the fields are now decoupled and have separate solutions where $\phi_3'$ is stable and $\phi_4'$ has both a ghost and a tachyon. This corresponds to a situation in which $\phi_4'$'s equation of motion has just been multiplied by a global minus sign, and is thus stable. Equivalently, we can understand this result from the Hamiltonian of this two field system where the parts that belong to each corresponding field have opposite energies. However, because the two fields no longer interact, they conserve their energy separately and we now effectively have two Hamiltonians with one bounded from above and the other bounded from below. 

We conclude that for restricted values of the coefficients of the initial interacting action, we can have a stable system despite the presence of two ghosts, and this happens because the terms with opposite energy contributions are decoupled from each other.

\subsubsection{High Energy} 
In the limit $k \to \infty$ we obtain the following kinetic and mass terms in the transformed action:
\begin{align}
& K_3'\rightarrow -\frac{1}{c_1^2}, \quad M_3^{'2}\rightarrow \frac{k^2}{c_1^2}, \nonumber \\ 
& K_4'\rightarrow -\frac{1}{c_1^2}, \quad M_4^{'2}\rightarrow  \frac{k^2}{c_1^2}. 
\end{align}
In this case, we find that the high-energy limit gives us an unstable solution with two ghosts. 
Here, we have effectively decoupled solutions that correspond to eq.~(\ref{two_ghost_limit}) when $D$ is not relevant and thus go as $e^{kt}$. 

\section{Discussion}\label{sec:discussion}

There are a number of cosmological models in the literature that have a dynamical evolution of dark energy, and provide alternatives to the standard cosmological constant paradigm. These models typically describe dark energy as one additional physical field that leads to the late-time accelerated expansion of the universe. One of the major challenges in constructing consistent dark energy models has been the fact that they are plagued by instabilities that appear in the evolution of linear cosmological perturbations, and may render these models phenomenologically unviable. 

In this paper, we generically study properties of instabilities in theories that have two interacting scalar cosmological fields as a proxy for dark energy and a matter field (and thus encompass scalar-tensor, vector-tensor and bigravity theories), although the arguments and results can be extended to more interacting fields as well. In particular, we have generically explored the relationship between three types of instabilities: ghosts (associated to negative kinetic energy of fields), gradients (associated to negative squared momenta) and tachyons (associated to negative squared masses). We have used canonical transformations to show explicitly that at the level of the action, there is an equivalence between ghost, gradient, and tachyonic instabilities in the low and high-energy regimes. 

At high energies, if any of the two fields exhibits a ghost or a gradient instability then the entire system becomes unstable, leading to a runaway growing behaviour of any fields interacting to the manifestly unstable field.
Similarly, at low energies, if any of the two fields exhibits a ghost or a tachyonic instability then the entire system becomes unstable. Intuitively, this happens because the total energy of the system is composed of positive and negative terms that are allowed to exchange energy. As a consequence, both the negative and positive terms can grow arbitrarily large while still keeping the total energy conserved, which leads to the aforementioned runaway behaviour in fields.

We thus conclude that high-energy ghosts and gradients in any field are all equivalent, and low-energy ghosts and tachyons are also equivalent. This result is found for interacting physical systems, to which we apply canonical transformations that allow us to bring any physical system into mathematical forms that make the relationship of these instabilities manifest. Via examples, we explicitly show that the presence of instabilities in an interacting system is a physical property that can manifest in different ways mathematically. Ultimately, the instability will be present regardless of whether it appears as negative kinetic energy or negative mass terms in a given field basis, and regardless of the specific field that exhibits the negative terms.

Understanding the relationship between these instabilities is essential in evaluating when a given theory can be phenomenologically viable. Typically, high-energy instabilities need to be avoided, and thus conditions on avoiding both high-energy ghosts and gradient instabilities must be imposed. Meanwhile, low-energy instabilities are allowed, and thus both kinetic and mass terms can be negative in this regime. Indeed, in cosmology, at low energies it is crucial to allow for a growth of perturbations as they lead to the formation of large-scale structures in the universe. 
Nevertheless, the timescale of this growth must be kept under control in order to comply with observations. 


In the case of interacting systems, we analyze the timescale of instabilities whenever they are present. We show that for most models, the presence of interactions will not alter whether an instability is present or not. Nevertheless, the value of this interaction term does affect the actual timescale of the instability in the fields. 
Therefore, in cosmology, while one may naively impose only conditions on the masses of fields to control the growth rate of low-energy instabilities, the appropriate approach may be to impose conditions on generalized quantities that take intro consideration interaction terms.


In addition, we show that there are rare models that may exhibit negative kinetic energy terms and hence are seemingly unstable, but the presence of interactions actually lead to stable solutions for all fields. We show that in this case it is possible to apply a canonical transformation to bring the system into a form where fields are effectively decoupled, and one of them has a positive-definite total energy, whereas the other has a negative-definite energy. In this case, both fields evolve independently and do not exchange energy, which renders the system stable. These models show that, again, it is important to analyze and appropriately consider interactions when searching for viable models. In particular, while high-energy instabilities are usually avoided by imposing positivity of all the coefficients in the action, it may be possible to relax these conditions and allow for negative terms if the presence of interactions help obtain stable solutions anyway.

As an example for how the results of this work can be used in practice, we mention the simplest case of scalar-tensor theories with second-order temporal and spatial derivative equations of motion. First of all, we must bring the system into the form of eq.~(\ref{twofields}) (this has been explicitly done in \cite{DeFelice:2016ucp}). Afterwards, we study the sign of the kinetic $K_{1,2}(t,k)$ and mass $m_{1,2}(t,k)$ terms in the low and high-energy regimes, while studying also the behaviour of the interaction term $D$. From the result obtained in \cite{DeFelice:2016ucp}, we find that in the high-energy regime $D\propto k^{-1}$, $K_{1,2}\propto k^0$, $m_{1,2}\propto k$, and hence $D$ becomes irrelevant in determining the fields evolution. Therefore, in order to avoid high-energy instabilities, we would simply have to  impose $(\lim_{k\to \infty} K_{1,2})>0$ and $(\lim_{k\to \infty} m^2_{1,2})>0$. On the other hand, in the low-energy regime, we find that $D\propto k$, $K_{1,2}\propto k^0$, $m_{1,2}\propto k^0$, and hence $D$ again becomes irrelevant in determining the evolution of the fields. In this case, we would relax the no ghost condition and impose conditions on the action coefficients that allow negative signs in either the kinetic or mass terms provided the timescale is of order of the Hubble rate or slower.
For these simple scalar-tensor theories we see that $D$ had no relevant role in the low and high-energy regimes. However, this may not necessarily be the case in models with a more complicated $k$-dependence of the kinetic and mass coefficients, such as in DHOST, vector-tensor, or bimetric models. Whether $D$ is relevant or not, must be ultimately determined case by case.  

In conclusion, we have shown that for many generic cosmological models that promote dark energy to a dynamical field, instabilities in the system may manifest themselves in different mathematical ways, although they are physically equivalent. We show this equivalence by applying canonical transformations
at the level of the action that describes linear scalar perturbations in these models. Furthermore, we have shown that interactions in the high and low-energy regimes may affect the stability of the system or the timescale of the instabilities when present. These results can aid in the study of these general cosmological models and inform the conditions we need to impose on these theories when searching for phenomenologically viable models.

\section*{Acknowledgments}
\vspace{-0.2in}
We are grateful to discussions with Pedro Ferreira, and Lian Tao Wang. WW is supported at the University of Chicago by the Physical Sciences Division and the Kavli Institute for Cosmological Physics. ML is supported at the University of Chicago by the Kavli Institute for Cosmological Physics through an endowment from the Kavli Foundation and its founder Fred Kavli.

\appendix*
\section{Two field action}

\subsection{Transformed Lagrangian}\label{App:twofieldcoeffs}
The transformed Lagrangian for two fields is given by:
\begin{align}\label{transfLtwofields2}
 \mathcal{L'} &= \frac{C_1}{2A}\Dot{\phi_1'}^2 + \frac{C_2}{A}\Dot{\phi_1'}\phi_1' - \frac{C_3}{2A} \phi_1'^2 + \frac{C_4}{2A}\Dot{\phi_2'}^2 + \frac{C_5}{A}\Dot{\phi_2'}\phi_2' 
    \nonumber\\ & -\frac{C_6}{2A} \phi_2'^2 + \frac{C_7}{A}\Dot{\phi_1'}\Dot{\phi_2'} +  \frac{C_8}{A}\Dot{\phi_1'}\phi_2' +  \frac{C_9}{A}\Dot{\phi_2'}\phi_1' +  \frac{C_{10}}{A}\phi_1'\phi_2'.
\end{align}

\begin{align}
A&\equiv a_2^2 \left(c_2^2 \left(D^2+k_2 m_1^2\right) \left(D^2+k_1 m_2^2\right)+c_1^2 k_1 m_1^2\right) \nonumber \\
&+2 a_1 a_2 c_1 c_2 D^2+a_1^2 \left(c_2^2 k_2 m_2^2+c_1^2\right).
\end{align}

Furthermore, the coefficients in front of the various terms are explicitly given by:
\begin{align}
C_1 &=c_2^2 k_2 \left(D^2+k_1 m_2^2\right)+c_1^2 k_1, \nonumber \\
C_2 &=\left[a_1 \left(b_2 c_1 c_2 D^2+b_1 \left(c_2^2 k_2 m_2^2+c_1^2\right)\right)\right. \nonumber\\
&\left.+a_2 \left(b_2 \left(c_2^2 \left(D^2+k_2 m_1^2\right) \left(D^2+k_1 m_2^2\right)+c_1^2 k_1 m_1^2\right)\right.\right. \nonumber \\
&\left.\left.+b_1 c_1 c_2 D^2\right)\right], \nonumber \\
C_3&=c_2^2 m_2^2 \left(D^2+k_2 m_1^2\right)+c_1^2 m_1^2, \nonumber \\
C_4&=a_2^2 k_1 \left(D^2+k_2 m_1^2\right)+a_1^2 k_2, \nonumber \\
C_5&=\left[a_1 a_2 D^2 \left(c_2 d_1+c_1 d_2\right)+a_1^2 \left(c_2 d_2 k_2 m_2^2+c_1 d_1\right)\right. \nonumber \\
&\left.+a_2^2 \left(c_2 d_2 \left(D^2+k_2 m_1^2\right) \left(D^2+k_1 m_2^2\right)+c_1 d_1 k_1 m_1^2\right)\right], \nonumber \\
C_6&=a_2^2 m_1^2 \left(D^2+k_1 m_2^2\right)+a_1^2 m_2^2, \nonumber \\
C_7&=D \left(a_2 c_1 k_1-a_1 c_2 k_2\right), \nonumber \\
C_8&=D \left(a_2 c_2 \left(D^2+k_1 m_2^2\right)+a_1 c_1\right), \nonumber \\
C_9&=-D \left(a_2 c_2 \left(D^2+k_2 m_1^2\right)+a_1 c_1\right), \nonumber \\
C_{10}&=D \left(a_2 c_1 m_1^2-a_1 c_2 m_2^2\right).
\end{align}

This transformed Lagrangian reduces to the single field result for both $\phi_1$ and $\phi_2$ when $D\to 0$ as expected.

\subsection{Interaction terms}\label{App:AntisymmetricTerms}
Consider terms in our Lagrangian of the following:
\begin{equation}
\int d^4x \left[D_1 \Dot{\phi_1}\phi_2 +D_2 \Dot{\phi_2}\phi_1 \right].
\end{equation}
They can be integrated by parts to bring them into an antisymmetric form as desired for the study of stability. We rewrite these terms as:
\begin{align}
&\int d^4x\left[  -\Dot{D_1} \phi_1\phi_2 -D_1 \Dot{\phi_2}\phi_1 +D_2 \Dot{\phi_2}\phi_1 \right] \nonumber
\\  = &\int d^4x\left[ -\Dot{D_1} \phi_1\phi_2 + (D_2-D_1) \Dot{\phi_2} \phi_1\right] .
\end{align}
After another integration by parts this reads:
\begin{align}
&\int d^4x\left[-\Dot{D_1} \phi_1\phi_2 + \frac{1}{2}(D_2-D_1) \Dot{\phi_2} \phi_1 + \frac{1}{2}(D_2-D_1) \Dot{\phi_2} \phi_1 \right] \nonumber
\\  \nonumber=&\int d^4x \left[-\Dot{D_1} \phi_1\phi_2 + \frac{1}{2}(D_2-D_1) \Dot{\phi_2} \phi_1-\frac{1}{2}(D_2-D_1) \Dot{\phi_1} \phi_2\right.
\\  &\left.-\frac{1}{2}(\Dot{D_2}-\Dot{D_1})\phi_1\phi_2\right]
\end{align}
This leads to the final expression:
\begin{align}
\int d^4x \left[ \frac{1}{2}(D_2-D_1)\left[\Dot{\phi_2}\phi_1-\Dot{\phi_1}\phi_2\right]-\frac{1}{2}(\Dot{D_2}-\Dot{D_1})\phi_1\phi_2 \right].
\end{align}
Here we can see that this procedure always leads to an antisymmetric term that will not affect our stability analysis as noted earlier, as well as a mass mixing term that depends on time. If our coefficients do not depend on time this last contribution will vanish.

\bibliography{RefModifiedGravity} 

\begin{thebibliography}{46}%
\makeatletter
\providecommand \@ifxundefined [1]{%
 \@ifx{#1\undefined}
}%
\providecommand \@ifnum [1]{%
 \ifnum #1\expandafter \@firstoftwo
 \else \expandafter \@secondoftwo
 \fi
}%
\providecommand \@ifx [1]{%
 \ifx #1\expandafter \@firstoftwo
 \else \expandafter \@secondoftwo
 \fi
}%
\providecommand \natexlab [1]{#1}%
\providecommand \enquote  [1]{``#1''}%
\providecommand \bibnamefont  [1]{#1}%
\providecommand \bibfnamefont [1]{#1}%
\providecommand \citenamefont [1]{#1}%
\providecommand \href@noop [0]{\@secondoftwo}%
\providecommand \href [0]{\begingroup \@sanitize@url \@href}%
\providecommand \@href[1]{\@@startlink{#1}\@@href}%
\providecommand \@@href[1]{\endgroup#1\@@endlink}%
\providecommand \@sanitize@url [0]{\catcode `\\12\catcode `\$12\catcode
  `\&12\catcode `\#12\catcode `\^12\catcode `\_12\catcode `\%12\relax}%
\providecommand \@@startlink[1]{}%
\providecommand \@@endlink[0]{}%
\providecommand \url  [0]{\begingroup\@sanitize@url \@url }%
\providecommand \@url [1]{\endgroup\@href {#1}{\urlprefix }}%
\providecommand \urlprefix  [0]{URL }%
\providecommand \Eprint [0]{\href }%
\providecommand \doibase [0]{http://dx.doi.org/}%
\providecommand \selectlanguage [0]{\@gobble}%
\providecommand \bibinfo  [0]{\@secondoftwo}%
\providecommand \bibfield  [0]{\@secondoftwo}%
\providecommand \translation [1]{[#1]}%
\providecommand \BibitemOpen [0]{}%
\providecommand \bibitemStop [0]{}%
\providecommand \bibitemNoStop [0]{.\EOS\space}%
\providecommand \EOS [0]{\spacefactor3000\relax}%
\providecommand \BibitemShut  [1]{\csname bibitem#1\endcsname}%
\let\auto@bib@innerbib\@empty
\bibitem [{\citenamefont {Aghanim}\ \emph {et~al.}(2018)\citenamefont {Aghanim}
  \emph {et~al.}}]{Aghanim:2018eyx}%
  \BibitemOpen
  \bibfield  {author} {\bibinfo {author} {\bibfnamefont {N.}~\bibnamefont
  {Aghanim}} \emph {et~al.} (\bibinfo {collaboration} {Planck}),\ }\href@noop
  {} {\  (\bibinfo {year} {2018})},\ \Eprint {http://arxiv.org/abs/1807.06209}
  {arXiv:1807.06209 [astro-ph.CO]} \BibitemShut {NoStop}%
\bibitem [{\citenamefont {Martin}(2012)}]{Martin:2012bt}%
  \BibitemOpen
  \bibfield  {author} {\bibinfo {author} {\bibfnamefont {J.}~\bibnamefont
  {Martin}},\ }\href {\doibase 10.1016/j.crhy.2012.04.008} {\bibfield
  {journal} {\bibinfo  {journal} {Comptes Rendus Physique}\ }\textbf {\bibinfo
  {volume} {13}},\ \bibinfo {pages} {566} (\bibinfo {year} {2012})},\ \Eprint
  {http://arxiv.org/abs/1205.3365} {arXiv:1205.3365 [astro-ph.CO]} \BibitemShut
  {NoStop}%
\bibitem [{\citenamefont {Clifton}\ \emph {et~al.}(2012)\citenamefont
  {Clifton}, \citenamefont {Ferreira}, \citenamefont {Padilla},\ and\
  \citenamefont {Skordis}}]{Clifton:2011jh}%
  \BibitemOpen
  \bibfield  {author} {\bibinfo {author} {\bibfnamefont {T.}~\bibnamefont
  {Clifton}}, \bibinfo {author} {\bibfnamefont {P.~G.}\ \bibnamefont
  {Ferreira}}, \bibinfo {author} {\bibfnamefont {A.}~\bibnamefont {Padilla}}, \
  and\ \bibinfo {author} {\bibfnamefont {C.}~\bibnamefont {Skordis}},\ }\href
  {\doibase 10.1016/j.physrep.2012.01.001} {\bibfield  {journal} {\bibinfo
  {journal} {Phys. Rept.}\ }\textbf {\bibinfo {volume} {513}},\ \bibinfo
  {pages} {1} (\bibinfo {year} {2012})},\ \Eprint
  {http://arxiv.org/abs/1106.2476} {arXiv:1106.2476 [astro-ph.CO]} \BibitemShut
  {NoStop}%
\bibitem [{\citenamefont {Joyce}\ \emph {et~al.}(2015)\citenamefont {Joyce},
  \citenamefont {Jain}, \citenamefont {Khoury},\ and\ \citenamefont
  {Trodden}}]{Joyce:2014kja}%
  \BibitemOpen
  \bibfield  {author} {\bibinfo {author} {\bibfnamefont {A.}~\bibnamefont
  {Joyce}}, \bibinfo {author} {\bibfnamefont {B.}~\bibnamefont {Jain}},
  \bibinfo {author} {\bibfnamefont {J.}~\bibnamefont {Khoury}}, \ and\ \bibinfo
  {author} {\bibfnamefont {M.}~\bibnamefont {Trodden}},\ }\href {\doibase
  10.1016/j.physrep.2014.12.002} {\bibfield  {journal} {\bibinfo  {journal}
  {Phys. Rept.}\ }\textbf {\bibinfo {volume} {568}},\ \bibinfo {pages} {1}
  (\bibinfo {year} {2015})},\ \Eprint {http://arxiv.org/abs/1407.0059}
  {arXiv:1407.0059 [astro-ph.CO]} \BibitemShut {NoStop}%
\bibitem [{\citenamefont {Joyce}\ \emph {et~al.}(2016)\citenamefont {Joyce},
  \citenamefont {Lombriser},\ and\ \citenamefont {Schmidt}}]{Joyce:2016vqv}%
  \BibitemOpen
  \bibfield  {author} {\bibinfo {author} {\bibfnamefont {A.}~\bibnamefont
  {Joyce}}, \bibinfo {author} {\bibfnamefont {L.}~\bibnamefont {Lombriser}}, \
  and\ \bibinfo {author} {\bibfnamefont {F.}~\bibnamefont {Schmidt}},\ }\href
  {\doibase 10.1146/annurev-nucl-102115-044553} {\bibfield  {journal} {\bibinfo
   {journal} {Ann. Rev. Nucl. Part. Sci.}\ }\textbf {\bibinfo {volume} {66}},\
  \bibinfo {pages} {95} (\bibinfo {year} {2016})},\ \Eprint
  {http://arxiv.org/abs/1601.06133} {arXiv:1601.06133 [astro-ph.CO]}
  \BibitemShut {NoStop}%
\bibitem [{\citenamefont {Ferreira}(2019)}]{Ferreira:2019xrr}%
  \BibitemOpen
  \bibfield  {author} {\bibinfo {author} {\bibfnamefont {P.~G.}\ \bibnamefont
  {Ferreira}},\ }\href@noop {} {\bibfield  {journal} {\bibinfo  {journal}
  {Annual Review of Astronomy and Astrophysics}\ }\textbf {\bibinfo {volume}
  {57}},\ \bibinfo {pages} {335–374} (\bibinfo {year} {2019})},\ \Eprint
  {http://arxiv.org/abs/1902.10503} {arXiv:1902.10503 [astro-ph.CO]}
  \BibitemShut {NoStop}%
\bibitem [{\citenamefont {Kodama}\ and\ \citenamefont
  {Sasaki}(1984)}]{Kodama:1985bj}%
  \BibitemOpen
  \bibfield  {author} {\bibinfo {author} {\bibfnamefont {H.}~\bibnamefont
  {Kodama}}\ and\ \bibinfo {author} {\bibfnamefont {M.}~\bibnamefont
  {Sasaki}},\ }\href {\doibase 10.1143/PTPS.78.1} {\bibfield  {journal}
  {\bibinfo  {journal} {Prog. Theor. Phys. Suppl.}\ }\textbf {\bibinfo {volume}
  {78}},\ \bibinfo {pages} {1} (\bibinfo {year} {1984})}\BibitemShut {NoStop}%
\bibitem [{\citenamefont {Horndeski}(1974)}]{Horndeski:1974wa}%
  \BibitemOpen
  \bibfield  {author} {\bibinfo {author} {\bibfnamefont {G.~W.}\ \bibnamefont
  {Horndeski}},\ }\href {\doibase 10.1007/BF01807638} {\bibfield  {journal}
  {\bibinfo  {journal} {Int. J. Theor. Phys.}\ }\textbf {\bibinfo {volume}
  {10}},\ \bibinfo {pages} {363} (\bibinfo {year} {1974})}\BibitemShut
  {NoStop}%
\bibitem [{\citenamefont {Deffayet}\ \emph {et~al.}(2009)\citenamefont
  {Deffayet}, \citenamefont {Esposito-Farese},\ and\ \citenamefont
  {Vikman}}]{Deffayet:2009wt}%
  \BibitemOpen
  \bibfield  {author} {\bibinfo {author} {\bibfnamefont {C.}~\bibnamefont
  {Deffayet}}, \bibinfo {author} {\bibfnamefont {G.}~\bibnamefont
  {Esposito-Farese}}, \ and\ \bibinfo {author} {\bibfnamefont {A.}~\bibnamefont
  {Vikman}},\ }\href {\doibase 10.1103/PhysRevD.79.084003} {\bibfield
  {journal} {\bibinfo  {journal} {Phys. Rev.}\ }\textbf {\bibinfo {volume}
  {D79}},\ \bibinfo {pages} {084003} (\bibinfo {year} {2009})},\ \Eprint
  {http://arxiv.org/abs/0901.1314} {arXiv:0901.1314 [hep-th]} \BibitemShut
  {NoStop}%
\bibitem [{\citenamefont {Langlois}\ and\ \citenamefont
  {Noui}(2016)}]{Langlois:2015cwa}%
  \BibitemOpen
  \bibfield  {author} {\bibinfo {author} {\bibfnamefont {D.}~\bibnamefont
  {Langlois}}\ and\ \bibinfo {author} {\bibfnamefont {K.}~\bibnamefont
  {Noui}},\ }\href {\doibase 10.1088/1475-7516/2016/02/034} {\bibfield
  {journal} {\bibinfo  {journal} {JCAP}\ }\textbf {\bibinfo {volume} {1602}},\
  \bibinfo {pages} {034} (\bibinfo {year} {2016})},\ \Eprint
  {http://arxiv.org/abs/1510.06930} {arXiv:1510.06930 [gr-qc]} \BibitemShut
  {NoStop}%
\bibitem [{\citenamefont {Ben~Achour}\ \emph {et~al.}(2016)\citenamefont
  {Ben~Achour}, \citenamefont {Crisostomi}, \citenamefont {Koyama},
  \citenamefont {Langlois}, \citenamefont {Noui},\ and\ \citenamefont
  {Tasinato}}]{BenAchour:2016fzp}%
  \BibitemOpen
  \bibfield  {author} {\bibinfo {author} {\bibfnamefont {J.}~\bibnamefont
  {Ben~Achour}}, \bibinfo {author} {\bibfnamefont {M.}~\bibnamefont
  {Crisostomi}}, \bibinfo {author} {\bibfnamefont {K.}~\bibnamefont {Koyama}},
  \bibinfo {author} {\bibfnamefont {D.}~\bibnamefont {Langlois}}, \bibinfo
  {author} {\bibfnamefont {K.}~\bibnamefont {Noui}}, \ and\ \bibinfo {author}
  {\bibfnamefont {G.}~\bibnamefont {Tasinato}},\ }\href {\doibase
  10.1007/JHEP12(2016)100} {\bibfield  {journal} {\bibinfo  {journal} {JHEP}\
  }\textbf {\bibinfo {volume} {12}},\ \bibinfo {pages} {100} (\bibinfo {year}
  {2016})},\ \Eprint {http://arxiv.org/abs/1608.08135} {arXiv:1608.08135
  [hep-th]} \BibitemShut {NoStop}%
\bibitem [{\citenamefont {Heisenberg}(2014)}]{Heisenberg:2014rta}%
  \BibitemOpen
  \bibfield  {author} {\bibinfo {author} {\bibfnamefont {L.}~\bibnamefont
  {Heisenberg}},\ }\href {\doibase 10.1088/1475-7516/2014/05/015} {\bibfield
  {journal} {\bibinfo  {journal} {JCAP}\ }\textbf {\bibinfo {volume} {1405}},\
  \bibinfo {pages} {015} (\bibinfo {year} {2014})},\ \Eprint
  {http://arxiv.org/abs/1402.7026} {arXiv:1402.7026 [hep-th]} \BibitemShut
  {NoStop}%
\bibitem [{\citenamefont {Zlosnik}\ \emph {et~al.}(2007)\citenamefont
  {Zlosnik}, \citenamefont {Ferreira},\ and\ \citenamefont
  {Starkman}}]{Zlosnik:2006zu}%
  \BibitemOpen
  \bibfield  {author} {\bibinfo {author} {\bibfnamefont {T.~G.}\ \bibnamefont
  {Zlosnik}}, \bibinfo {author} {\bibfnamefont {P.~G.}\ \bibnamefont
  {Ferreira}}, \ and\ \bibinfo {author} {\bibfnamefont {G.~D.}\ \bibnamefont
  {Starkman}},\ }\href {\doibase 10.1103/PhysRevD.75.044017} {\bibfield
  {journal} {\bibinfo  {journal} {Phys. Rev.}\ }\textbf {\bibinfo {volume}
  {D75}},\ \bibinfo {pages} {044017} (\bibinfo {year} {2007})},\ \Eprint
  {http://arxiv.org/abs/astro-ph/0607411} {arXiv:astro-ph/0607411 [astro-ph]}
  \BibitemShut {NoStop}%
\bibitem [{\citenamefont {de~Rham}\ \emph {et~al.}(2011)\citenamefont
  {de~Rham}, \citenamefont {Gabadadze},\ and\ \citenamefont
  {Tolley}}]{deRham:2010kj}%
  \BibitemOpen
  \bibfield  {author} {\bibinfo {author} {\bibfnamefont {C.}~\bibnamefont
  {de~Rham}}, \bibinfo {author} {\bibfnamefont {G.}~\bibnamefont {Gabadadze}},
  \ and\ \bibinfo {author} {\bibfnamefont {A.~J.}\ \bibnamefont {Tolley}},\
  }\href {\doibase 10.1103/PhysRevLett.106.231101} {\bibfield  {journal}
  {\bibinfo  {journal} {Phys.Rev.Lett.}\ }\textbf {\bibinfo {volume} {106}},\
  \bibinfo {pages} {231101} (\bibinfo {year} {2011})},\ \Eprint
  {http://arxiv.org/abs/1011.1232} {arXiv:1011.1232 [hep-th]} \BibitemShut
  {NoStop}%
\bibitem [{\citenamefont {Hassan}\ and\ \citenamefont
  {Rosen}(2012)}]{Hassan:2011zd}%
  \BibitemOpen
  \bibfield  {author} {\bibinfo {author} {\bibfnamefont {S.}~\bibnamefont
  {Hassan}}\ and\ \bibinfo {author} {\bibfnamefont {R.~A.}\ \bibnamefont
  {Rosen}},\ }\href {\doibase 10.1007/JHEP02(2012)126} {\bibfield  {journal}
  {\bibinfo  {journal} {JHEP}\ }\textbf {\bibinfo {volume} {1202}},\ \bibinfo
  {pages} {126} (\bibinfo {year} {2012})},\ \Eprint
  {http://arxiv.org/abs/1109.3515} {arXiv:1109.3515 [hep-th]} \BibitemShut
  {NoStop}%
\bibitem [{\citenamefont {Arkani-Hamed}\ \emph {et~al.}(2004)\citenamefont
  {Arkani-Hamed}, \citenamefont {Cheng}, \citenamefont {Luty},\ and\
  \citenamefont {Mukohyama}}]{ArkaniHamed:2003uy}%
  \BibitemOpen
  \bibfield  {author} {\bibinfo {author} {\bibfnamefont {N.}~\bibnamefont
  {Arkani-Hamed}}, \bibinfo {author} {\bibfnamefont {H.-C.}\ \bibnamefont
  {Cheng}}, \bibinfo {author} {\bibfnamefont {M.~A.}\ \bibnamefont {Luty}}, \
  and\ \bibinfo {author} {\bibfnamefont {S.}~\bibnamefont {Mukohyama}},\ }\href
  {\doibase 10.1088/1126-6708/2004/05/074} {\bibfield  {journal} {\bibinfo
  {journal} {JHEP}\ }\textbf {\bibinfo {volume} {05}},\ \bibinfo {pages} {074}
  (\bibinfo {year} {2004})},\ \Eprint {http://arxiv.org/abs/hep-th/0312099}
  {arXiv:hep-th/0312099 [hep-th]} \BibitemShut {NoStop}%
\bibitem [{\citenamefont {Izumi}\ and\ \citenamefont
  {Tanaka}(2009)}]{Izumi:2007pb}%
  \BibitemOpen
  \bibfield  {author} {\bibinfo {author} {\bibfnamefont {K.}~\bibnamefont
  {Izumi}}\ and\ \bibinfo {author} {\bibfnamefont {T.}~\bibnamefont {Tanaka}},\
  }\href {\doibase 10.1143/PTP.121.419} {\bibfield  {journal} {\bibinfo
  {journal} {Prog. Theor. Phys.}\ }\textbf {\bibinfo {volume} {121}},\ \bibinfo
  {pages} {419} (\bibinfo {year} {2009})},\ \Eprint
  {http://arxiv.org/abs/0709.0199} {arXiv:0709.0199 [gr-qc]} \BibitemShut
  {NoStop}%
\bibitem [{\citenamefont {Garriga}\ and\ \citenamefont
  {Vilenkin}(2013)}]{Garriga:2012pk}%
  \BibitemOpen
  \bibfield  {author} {\bibinfo {author} {\bibfnamefont {J.}~\bibnamefont
  {Garriga}}\ and\ \bibinfo {author} {\bibfnamefont {A.}~\bibnamefont
  {Vilenkin}},\ }\href {\doibase 10.1088/1475-7516/2013/01/036} {\bibfield
  {journal} {\bibinfo  {journal} {JCAP}\ }\textbf {\bibinfo {volume} {1301}},\
  \bibinfo {pages} {036} (\bibinfo {year} {2013})},\ \Eprint
  {http://arxiv.org/abs/1202.1239} {arXiv:1202.1239 [hep-th]} \BibitemShut
  {NoStop}%
\bibitem [{\citenamefont {Sbisa}(2015)}]{Sbisa:2014pzo}%
  \BibitemOpen
  \bibfield  {author} {\bibinfo {author} {\bibfnamefont {F.}~\bibnamefont
  {Sbisa}},\ }\href {\doibase 10.1088/0143-0807/36/1/015009} {\bibfield
  {journal} {\bibinfo  {journal} {Eur. J. Phys.}\ }\textbf {\bibinfo {volume}
  {36}},\ \bibinfo {pages} {015009} (\bibinfo {year} {2015})},\ \Eprint
  {http://arxiv.org/abs/1406.4550} {arXiv:1406.4550 [hep-th]} \BibitemShut
  {NoStop}%
\bibitem [{\citenamefont {Koennig}\ \emph {et~al.}(2016)\citenamefont
  {Koennig}, \citenamefont {Nersisyan}, \citenamefont {Akrami}, \citenamefont
  {Amendola},\ and\ \citenamefont {Zumalacárregui}}]{Konnig:2016idp}%
  \BibitemOpen
  \bibfield  {author} {\bibinfo {author} {\bibfnamefont {F.}~\bibnamefont
  {Koennig}}, \bibinfo {author} {\bibfnamefont {H.}~\bibnamefont {Nersisyan}},
  \bibinfo {author} {\bibfnamefont {Y.}~\bibnamefont {Akrami}}, \bibinfo
  {author} {\bibfnamefont {L.}~\bibnamefont {Amendola}}, \ and\ \bibinfo
  {author} {\bibfnamefont {M.}~\bibnamefont {Zumalacárregui}},\ }\href
  {\doibase 10.1007/JHEP11(2016)118} {\bibfield  {journal} {\bibinfo  {journal}
  {JHEP}\ }\textbf {\bibinfo {volume} {11}},\ \bibinfo {pages} {118} (\bibinfo
  {year} {2016})},\ \Eprint {http://arxiv.org/abs/1605.08757} {arXiv:1605.08757
  [gr-qc]} \BibitemShut {NoStop}%
\bibitem [{\citenamefont {Cheung}\ \emph {et~al.}(2008)\citenamefont {Cheung},
  \citenamefont {Creminelli}, \citenamefont {Fitzpatrick}, \citenamefont
  {Kaplan},\ and\ \citenamefont {Senatore}}]{Cheung:2007st}%
  \BibitemOpen
  \bibfield  {author} {\bibinfo {author} {\bibfnamefont {C.}~\bibnamefont
  {Cheung}}, \bibinfo {author} {\bibfnamefont {P.}~\bibnamefont {Creminelli}},
  \bibinfo {author} {\bibfnamefont {A.~L.}\ \bibnamefont {Fitzpatrick}},
  \bibinfo {author} {\bibfnamefont {J.}~\bibnamefont {Kaplan}}, \ and\ \bibinfo
  {author} {\bibfnamefont {L.}~\bibnamefont {Senatore}},\ }\href {\doibase
  10.1088/1126-6708/2008/03/014} {\bibfield  {journal} {\bibinfo  {journal}
  {JHEP}\ }\textbf {\bibinfo {volume} {03}},\ \bibinfo {pages} {014} (\bibinfo
  {year} {2008})},\ \Eprint {http://arxiv.org/abs/0709.0293} {arXiv:0709.0293
  [hep-th]} \BibitemShut {NoStop}%
\bibitem [{\citenamefont {Creminelli}\ \emph {et~al.}(2009)\citenamefont
  {Creminelli}, \citenamefont {D'Amico}, \citenamefont {Norena},\ and\
  \citenamefont {Vernizzi}}]{Creminelli:2008wc}%
  \BibitemOpen
  \bibfield  {author} {\bibinfo {author} {\bibfnamefont {P.}~\bibnamefont
  {Creminelli}}, \bibinfo {author} {\bibfnamefont {G.}~\bibnamefont {D'Amico}},
  \bibinfo {author} {\bibfnamefont {J.}~\bibnamefont {Norena}}, \ and\ \bibinfo
  {author} {\bibfnamefont {F.}~\bibnamefont {Vernizzi}},\ }\href {\doibase
  10.1088/1475-7516/2009/02/018} {\bibfield  {journal} {\bibinfo  {journal}
  {JCAP}\ }\textbf {\bibinfo {volume} {0902}},\ \bibinfo {pages} {018}
  (\bibinfo {year} {2009})},\ \Eprint {http://arxiv.org/abs/0811.0827}
  {arXiv:0811.0827 [astro-ph]} \BibitemShut {NoStop}%
\bibitem [{\citenamefont {De~Felice}\ and\ \citenamefont
  {Tsujikawa}(2012)}]{DeFelice:2011bh}%
  \BibitemOpen
  \bibfield  {author} {\bibinfo {author} {\bibfnamefont {A.}~\bibnamefont
  {De~Felice}}\ and\ \bibinfo {author} {\bibfnamefont {S.}~\bibnamefont
  {Tsujikawa}},\ }\href {\doibase 10.1088/1475-7516/2012/02/007} {\bibfield
  {journal} {\bibinfo  {journal} {JCAP}\ }\textbf {\bibinfo {volume} {1202}},\
  \bibinfo {pages} {007} (\bibinfo {year} {2012})},\ \Eprint
  {http://arxiv.org/abs/1110.3878} {arXiv:1110.3878 [gr-qc]} \BibitemShut
  {NoStop}%
\bibitem [{\citenamefont {Gubitosi}\ \emph {et~al.}(2013)\citenamefont
  {Gubitosi}, \citenamefont {Piazza},\ and\ \citenamefont
  {Vernizzi}}]{Gubitosi:2012hu}%
  \BibitemOpen
  \bibfield  {author} {\bibinfo {author} {\bibfnamefont {G.}~\bibnamefont
  {Gubitosi}}, \bibinfo {author} {\bibfnamefont {F.}~\bibnamefont {Piazza}}, \
  and\ \bibinfo {author} {\bibfnamefont {F.}~\bibnamefont {Vernizzi}},\ }\href
  {\doibase 10.1088/1475-7516/2013/02/032} {\bibfield  {journal} {\bibinfo
  {journal} {JCAP}\ }\textbf {\bibinfo {volume} {1302}},\ \bibinfo {pages}
  {032} (\bibinfo {year} {2013})},\ \bibinfo {note} {[JCAP1302,032(2013)]},\
  \Eprint {http://arxiv.org/abs/1210.0201} {arXiv:1210.0201 [hep-th]}
  \BibitemShut {NoStop}%
\bibitem [{\citenamefont {Bloomfield}\ \emph {et~al.}(2013)\citenamefont
  {Bloomfield}, \citenamefont {Flanagan}, \citenamefont {Park},\ and\
  \citenamefont {Watson}}]{Bloomfield:2012ff}%
  \BibitemOpen
  \bibfield  {author} {\bibinfo {author} {\bibfnamefont {J.~K.}\ \bibnamefont
  {Bloomfield}}, \bibinfo {author} {\bibfnamefont {E.~E.}\ \bibnamefont
  {Flanagan}}, \bibinfo {author} {\bibfnamefont {M.}~\bibnamefont {Park}}, \
  and\ \bibinfo {author} {\bibfnamefont {S.}~\bibnamefont {Watson}},\ }\href
  {\doibase 10.1088/1475-7516/2013/08/010} {\bibfield  {journal} {\bibinfo
  {journal} {JCAP}\ }\textbf {\bibinfo {volume} {1308}},\ \bibinfo {pages}
  {010} (\bibinfo {year} {2013})},\ \Eprint {http://arxiv.org/abs/1211.7054}
  {arXiv:1211.7054 [astro-ph.CO]} \BibitemShut {NoStop}%
\bibitem [{\citenamefont {Gleyzes}\ \emph {et~al.}(2013)\citenamefont
  {Gleyzes}, \citenamefont {Langlois}, \citenamefont {Piazza},\ and\
  \citenamefont {Vernizzi}}]{Gleyzes:2013ooa}%
  \BibitemOpen
  \bibfield  {author} {\bibinfo {author} {\bibfnamefont {J.}~\bibnamefont
  {Gleyzes}}, \bibinfo {author} {\bibfnamefont {D.}~\bibnamefont {Langlois}},
  \bibinfo {author} {\bibfnamefont {F.}~\bibnamefont {Piazza}}, \ and\ \bibinfo
  {author} {\bibfnamefont {F.}~\bibnamefont {Vernizzi}},\ }\href {\doibase
  10.1088/1475-7516/2013/08/025} {\bibfield  {journal} {\bibinfo  {journal}
  {JCAP}\ }\textbf {\bibinfo {volume} {1308}},\ \bibinfo {pages} {025}
  (\bibinfo {year} {2013})},\ \Eprint {http://arxiv.org/abs/1304.4840}
  {arXiv:1304.4840 [hep-th]} \BibitemShut {NoStop}%
\bibitem [{\citenamefont {Piazza}\ \emph {et~al.}(2014)\citenamefont {Piazza},
  \citenamefont {Steigerwald},\ and\ \citenamefont
  {Marinoni}}]{Piazza:2013pua}%
  \BibitemOpen
  \bibfield  {author} {\bibinfo {author} {\bibfnamefont {F.}~\bibnamefont
  {Piazza}}, \bibinfo {author} {\bibfnamefont {H.}~\bibnamefont {Steigerwald}},
  \ and\ \bibinfo {author} {\bibfnamefont {C.}~\bibnamefont {Marinoni}},\
  }\href {\doibase 10.1088/1475-7516/2014/05/043} {\bibfield  {journal}
  {\bibinfo  {journal} {JCAP}\ }\textbf {\bibinfo {volume} {1405}},\ \bibinfo
  {pages} {043} (\bibinfo {year} {2014})},\ \Eprint
  {http://arxiv.org/abs/1312.6111} {arXiv:1312.6111 [astro-ph.CO]} \BibitemShut
  {NoStop}%
\bibitem [{\citenamefont {Kase}\ and\ \citenamefont
  {Tsujikawa}(2015)}]{Kase:2014cwa}%
  \BibitemOpen
  \bibfield  {author} {\bibinfo {author} {\bibfnamefont {R.}~\bibnamefont
  {Kase}}\ and\ \bibinfo {author} {\bibfnamefont {S.}~\bibnamefont
  {Tsujikawa}},\ }\href {\doibase 10.1142/S0218271814430081} {\bibfield
  {journal} {\bibinfo  {journal} {Int. J. Mod. Phys.}\ }\textbf {\bibinfo
  {volume} {D23}},\ \bibinfo {pages} {1443008} (\bibinfo {year} {2015})},\
  \Eprint {http://arxiv.org/abs/1409.1984} {arXiv:1409.1984 [hep-th]}
  \BibitemShut {NoStop}%
\bibitem [{\citenamefont {Koennig}\ \emph {et~al.}(2014)\citenamefont
  {Koennig}, \citenamefont {Akrami}, \citenamefont {Amendola}, \citenamefont
  {Motta},\ and\ \citenamefont {Solomon}}]{Konnig:2014xva}%
  \BibitemOpen
  \bibfield  {author} {\bibinfo {author} {\bibfnamefont {F.}~\bibnamefont
  {Koennig}}, \bibinfo {author} {\bibfnamefont {Y.}~\bibnamefont {Akrami}},
  \bibinfo {author} {\bibfnamefont {L.}~\bibnamefont {Amendola}}, \bibinfo
  {author} {\bibfnamefont {M.}~\bibnamefont {Motta}}, \ and\ \bibinfo {author}
  {\bibfnamefont {A.~R.}\ \bibnamefont {Solomon}},\ }\href {\doibase
  10.1103/PhysRevD.90.124014} {\bibfield  {journal} {\bibinfo  {journal} {Phys.
  Rev.}\ }\textbf {\bibinfo {volume} {D90}},\ \bibinfo {pages} {124014}
  (\bibinfo {year} {2014})},\ \Eprint {http://arxiv.org/abs/1407.4331}
  {arXiv:1407.4331 [astro-ph.CO]} \BibitemShut {NoStop}%
\bibitem [{\citenamefont {Gleyzes}\ \emph {et~al.}(2015)\citenamefont
  {Gleyzes}, \citenamefont {Langlois}, \citenamefont {Mancarella},\ and\
  \citenamefont {Vernizzi}}]{Gleyzes:2015pma}%
  \BibitemOpen
  \bibfield  {author} {\bibinfo {author} {\bibfnamefont {J.}~\bibnamefont
  {Gleyzes}}, \bibinfo {author} {\bibfnamefont {D.}~\bibnamefont {Langlois}},
  \bibinfo {author} {\bibfnamefont {M.}~\bibnamefont {Mancarella}}, \ and\
  \bibinfo {author} {\bibfnamefont {F.}~\bibnamefont {Vernizzi}},\ }\href
  {\doibase 10.1088/1475-7516/2015/08/054} {\bibfield  {journal} {\bibinfo
  {journal} {JCAP}\ }\textbf {\bibinfo {volume} {1508}},\ \bibinfo {pages}
  {054} (\bibinfo {year} {2015})},\ \Eprint {http://arxiv.org/abs/1504.05481}
  {arXiv:1504.05481 [astro-ph.CO]} \BibitemShut {NoStop}%
\bibitem [{\citenamefont {Koennig}(2015)}]{Konnig:2015lfa}%
  \BibitemOpen
  \bibfield  {author} {\bibinfo {author} {\bibfnamefont {F.}~\bibnamefont
  {Koennig}},\ }\href {\doibase 10.1103/PhysRevD.91.104019} {\bibfield
  {journal} {\bibinfo  {journal} {Phys. Rev.}\ }\textbf {\bibinfo {volume}
  {D91}},\ \bibinfo {pages} {104019} (\bibinfo {year} {2015})},\ \Eprint
  {http://arxiv.org/abs/1503.07436} {arXiv:1503.07436 [astro-ph.CO]}
  \BibitemShut {NoStop}%
\bibitem [{\citenamefont {De~Felice}\ \emph {et~al.}(2016)\citenamefont
  {De~Felice}, \citenamefont {Heisenberg}, \citenamefont {Kase}, \citenamefont
  {Mukohyama}, \citenamefont {Tsujikawa},\ and\ \citenamefont
  {Zhang}}]{DeFelice:2016uil}%
  \BibitemOpen
  \bibfield  {author} {\bibinfo {author} {\bibfnamefont {A.}~\bibnamefont
  {De~Felice}}, \bibinfo {author} {\bibfnamefont {L.}~\bibnamefont
  {Heisenberg}}, \bibinfo {author} {\bibfnamefont {R.}~\bibnamefont {Kase}},
  \bibinfo {author} {\bibfnamefont {S.}~\bibnamefont {Mukohyama}}, \bibinfo
  {author} {\bibfnamefont {S.}~\bibnamefont {Tsujikawa}}, \ and\ \bibinfo
  {author} {\bibfnamefont {Y.-l.}\ \bibnamefont {Zhang}},\ }\href {\doibase
  10.1103/PhysRevD.94.044024} {\bibfield  {journal} {\bibinfo  {journal} {Phys.
  Rev.}\ }\textbf {\bibinfo {volume} {D94}},\ \bibinfo {pages} {044024}
  (\bibinfo {year} {2016})},\ \Eprint {http://arxiv.org/abs/1605.05066}
  {arXiv:1605.05066 [gr-qc]} \BibitemShut {NoStop}%
\bibitem [{\citenamefont {De~Felice}\ \emph
  {et~al.}(2017{\natexlab{a}})\citenamefont {De~Felice}, \citenamefont
  {Frusciante},\ and\ \citenamefont {Papadomanolakis}}]{DeFelice:2016ucp}%
  \BibitemOpen
  \bibfield  {author} {\bibinfo {author} {\bibfnamefont {A.}~\bibnamefont
  {De~Felice}}, \bibinfo {author} {\bibfnamefont {N.}~\bibnamefont
  {Frusciante}}, \ and\ \bibinfo {author} {\bibfnamefont {G.}~\bibnamefont
  {Papadomanolakis}},\ }\href {\doibase 10.1088/1475-7516/2017/03/027}
  {\bibfield  {journal} {\bibinfo  {journal} {JCAP}\ }\textbf {\bibinfo
  {volume} {1703}},\ \bibinfo {pages} {027} (\bibinfo {year}
  {2017}{\natexlab{a}})},\ \Eprint {http://arxiv.org/abs/1609.03599}
  {arXiv:1609.03599 [gr-qc]} \BibitemShut {NoStop}%
\bibitem [{\citenamefont {D'Amico}\ \emph {et~al.}(2017)\citenamefont
  {D'Amico}, \citenamefont {Huang}, \citenamefont {Mancarella},\ and\
  \citenamefont {Vernizzi}}]{DAmico:2016ntq}%
  \BibitemOpen
  \bibfield  {author} {\bibinfo {author} {\bibfnamefont {G.}~\bibnamefont
  {D'Amico}}, \bibinfo {author} {\bibfnamefont {Z.}~\bibnamefont {Huang}},
  \bibinfo {author} {\bibfnamefont {M.}~\bibnamefont {Mancarella}}, \ and\
  \bibinfo {author} {\bibfnamefont {F.}~\bibnamefont {Vernizzi}},\ }\href
  {\doibase 10.1088/1475-7516/2017/02/014} {\bibfield  {journal} {\bibinfo
  {journal} {JCAP}\ }\textbf {\bibinfo {volume} {1702}},\ \bibinfo {pages}
  {014} (\bibinfo {year} {2017})},\ \Eprint {http://arxiv.org/abs/1609.01272}
  {arXiv:1609.01272 [astro-ph.CO]} \BibitemShut {NoStop}%
\bibitem [{\citenamefont {Frusciante}\ \emph {et~al.}(2016)\citenamefont
  {Frusciante}, \citenamefont {Papadomanolakis},\ and\ \citenamefont
  {Silvestri}}]{Frusciante:2016xoj}%
  \BibitemOpen
  \bibfield  {author} {\bibinfo {author} {\bibfnamefont {N.}~\bibnamefont
  {Frusciante}}, \bibinfo {author} {\bibfnamefont {G.}~\bibnamefont
  {Papadomanolakis}}, \ and\ \bibinfo {author} {\bibfnamefont {A.}~\bibnamefont
  {Silvestri}},\ }\href {\doibase 10.1088/1475-7516/2016/07/018} {\bibfield
  {journal} {\bibinfo  {journal} {JCAP}\ }\textbf {\bibinfo {volume} {1607}},\
  \bibinfo {pages} {018} (\bibinfo {year} {2016})},\ \Eprint
  {http://arxiv.org/abs/1601.04064} {arXiv:1601.04064 [gr-qc]} \BibitemShut
  {NoStop}%
\bibitem [{\citenamefont {De~Felice}\ \emph
  {et~al.}(2017{\natexlab{b}})\citenamefont {De~Felice}, \citenamefont
  {Mukohyama},\ and\ \citenamefont {Oliosi}}]{DeFelice:2017wel}%
  \BibitemOpen
  \bibfield  {author} {\bibinfo {author} {\bibfnamefont {A.}~\bibnamefont
  {De~Felice}}, \bibinfo {author} {\bibfnamefont {S.}~\bibnamefont
  {Mukohyama}}, \ and\ \bibinfo {author} {\bibfnamefont {M.}~\bibnamefont
  {Oliosi}},\ }\href {\doibase 10.1103/PhysRevD.96.024032} {\bibfield
  {journal} {\bibinfo  {journal} {Phys. Rev.}\ }\textbf {\bibinfo {volume}
  {D96}},\ \bibinfo {pages} {024032} (\bibinfo {year} {2017}{\natexlab{b}})},\
  \Eprint {http://arxiv.org/abs/1701.01581} {arXiv:1701.01581 [hep-th]}
  \BibitemShut {NoStop}%
\bibitem [{\citenamefont {Lagos}\ \emph {et~al.}(2018)\citenamefont {Lagos},
  \citenamefont {Bellini}, \citenamefont {Noller}, \citenamefont {Ferreira},\
  and\ \citenamefont {Baker}}]{Lagos:2017hdr}%
  \BibitemOpen
  \bibfield  {author} {\bibinfo {author} {\bibfnamefont {M.}~\bibnamefont
  {Lagos}}, \bibinfo {author} {\bibfnamefont {E.}~\bibnamefont {Bellini}},
  \bibinfo {author} {\bibfnamefont {J.}~\bibnamefont {Noller}}, \bibinfo
  {author} {\bibfnamefont {P.~G.}\ \bibnamefont {Ferreira}}, \ and\ \bibinfo
  {author} {\bibfnamefont {T.}~\bibnamefont {Baker}},\ }\href {\doibase
  10.1088/1475-7516/2018/03/021} {\bibfield  {journal} {\bibinfo  {journal}
  {JCAP}\ }\textbf {\bibinfo {volume} {1803}},\ \bibinfo {pages} {021}
  (\bibinfo {year} {2018})},\ \Eprint {http://arxiv.org/abs/1711.09893}
  {arXiv:1711.09893 [gr-qc]} \BibitemShut {NoStop}%
\bibitem [{\citenamefont {Heisenberg}\ \emph {et~al.}(2018)\citenamefont
  {Heisenberg}, \citenamefont {Kase},\ and\ \citenamefont
  {Tsujikawa}}]{Heisenberg:2018mxx}%
  \BibitemOpen
  \bibfield  {author} {\bibinfo {author} {\bibfnamefont {L.}~\bibnamefont
  {Heisenberg}}, \bibinfo {author} {\bibfnamefont {R.}~\bibnamefont {Kase}}, \
  and\ \bibinfo {author} {\bibfnamefont {S.}~\bibnamefont {Tsujikawa}},\ }\href
  {\doibase 10.1103/PhysRevD.98.024038} {\bibfield  {journal} {\bibinfo
  {journal} {Phys. Rev.}\ }\textbf {\bibinfo {volume} {D98}},\ \bibinfo {pages}
  {024038} (\bibinfo {year} {2018})},\ \Eprint
  {http://arxiv.org/abs/1805.01066} {arXiv:1805.01066 [gr-qc]} \BibitemShut
  {NoStop}%
\bibitem [{\citenamefont {Frusciante}\ \emph {et~al.}(2019)\citenamefont
  {Frusciante}, \citenamefont {Papadomanolakis}, \citenamefont {Peirone},\ and\
  \citenamefont {Silvestri}}]{Frusciante:2018vht}%
  \BibitemOpen
  \bibfield  {author} {\bibinfo {author} {\bibfnamefont {N.}~\bibnamefont
  {Frusciante}}, \bibinfo {author} {\bibfnamefont {G.}~\bibnamefont
  {Papadomanolakis}}, \bibinfo {author} {\bibfnamefont {S.}~\bibnamefont
  {Peirone}}, \ and\ \bibinfo {author} {\bibfnamefont {A.}~\bibnamefont
  {Silvestri}},\ }\href {\doibase 10.1088/1475-7516/2019/02/029} {\bibfield
  {journal} {\bibinfo  {journal} {JCAP}\ }\textbf {\bibinfo {volume} {1902}},\
  \bibinfo {pages} {029} (\bibinfo {year} {2019})},\ \Eprint
  {http://arxiv.org/abs/1810.03461} {arXiv:1810.03461 [gr-qc]} \BibitemShut
  {NoStop}%
\bibitem [{\citenamefont {Crisostomi}\ \emph {et~al.}(2019)\citenamefont
  {Crisostomi}, \citenamefont {Koyama}, \citenamefont {Langlois}, \citenamefont
  {Noui},\ and\ \citenamefont {Steer}}]{Crisostomi:2018bsp}%
  \BibitemOpen
  \bibfield  {author} {\bibinfo {author} {\bibfnamefont {M.}~\bibnamefont
  {Crisostomi}}, \bibinfo {author} {\bibfnamefont {K.}~\bibnamefont {Koyama}},
  \bibinfo {author} {\bibfnamefont {D.}~\bibnamefont {Langlois}}, \bibinfo
  {author} {\bibfnamefont {K.}~\bibnamefont {Noui}}, \ and\ \bibinfo {author}
  {\bibfnamefont {D.~A.}\ \bibnamefont {Steer}},\ }\href {\doibase
  10.1088/1475-7516/2019/01/030} {\bibfield  {journal} {\bibinfo  {journal}
  {JCAP}\ }\textbf {\bibinfo {volume} {1901}},\ \bibinfo {pages} {030}
  (\bibinfo {year} {2019})},\ \Eprint {http://arxiv.org/abs/1810.12070}
  {arXiv:1810.12070 [hep-th]} \BibitemShut {NoStop}%
\bibitem [{\citenamefont {Frusciante}\ and\ \citenamefont
  {Perenon}(2019)}]{Frusciante:2019xia}%
  \BibitemOpen
  \bibfield  {author} {\bibinfo {author} {\bibfnamefont {N.}~\bibnamefont
  {Frusciante}}\ and\ \bibinfo {author} {\bibfnamefont {L.}~\bibnamefont
  {Perenon}},\ }\href@noop {} {\  (\bibinfo {year} {2019})},\ \Eprint
  {http://arxiv.org/abs/1907.03150} {arXiv:1907.03150 [astro-ph.CO]}
  \BibitemShut {NoStop}%
\bibitem [{\citenamefont {{Hu}}\ \emph {et~al.}(2014)\citenamefont {{Hu}},
  \citenamefont {{Raveri}}, \citenamefont {{Frusciante}},\ and\ \citenamefont
  {{Silvestri}}}]{EFTCAMB1}%
  \BibitemOpen
  \bibfield  {author} {\bibinfo {author} {\bibfnamefont {B.}~\bibnamefont
  {{Hu}}}, \bibinfo {author} {\bibfnamefont {M.}~\bibnamefont {{Raveri}}},
  \bibinfo {author} {\bibfnamefont {N.}~\bibnamefont {{Frusciante}}}, \ and\
  \bibinfo {author} {\bibfnamefont {A.}~\bibnamefont {{Silvestri}}},\ }\href
  {\doibase 10.1103/PhysRevD.89.103530} {\bibfield  {journal} {\bibinfo
  {journal} {Phys. Rev. D}\ }\textbf {\bibinfo {volume} {89}},\ \bibinfo {eid}
  {103530} (\bibinfo {year} {2014})},\ \Eprint {http://arxiv.org/abs/1312.5742}
  {arXiv:1312.5742} \BibitemShut {NoStop}%
\bibitem [{\citenamefont {Zumalacarregui}\ \emph {et~al.}(2017)\citenamefont
  {Zumalacarregui}, \citenamefont {Bellini}, \citenamefont {Sawicki},
  \citenamefont {Lesgourgues},\ and\ \citenamefont
  {Ferreira}}]{Zumalacarregui:2016pph}%
  \BibitemOpen
  \bibfield  {author} {\bibinfo {author} {\bibfnamefont {M.}~\bibnamefont
  {Zumalacarregui}}, \bibinfo {author} {\bibfnamefont {E.}~\bibnamefont
  {Bellini}}, \bibinfo {author} {\bibfnamefont {I.}~\bibnamefont {Sawicki}},
  \bibinfo {author} {\bibfnamefont {J.}~\bibnamefont {Lesgourgues}}, \ and\
  \bibinfo {author} {\bibfnamefont {P.~G.}\ \bibnamefont {Ferreira}},\ }\href
  {\doibase 10.1088/1475-7516/2017/08/019} {\bibfield  {journal} {\bibinfo
  {journal} {JCAP}\ }\textbf {\bibinfo {volume} {1708}},\ \bibinfo {pages}
  {019} (\bibinfo {year} {2017})},\ \Eprint {http://arxiv.org/abs/1605.06102}
  {arXiv:1605.06102 [astro-ph.CO]} \BibitemShut {NoStop}%
\bibitem [{\citenamefont {Gumrukcuoglu}\ \emph {et~al.}(2016)\citenamefont
  {Gumrukcuoglu}, \citenamefont {Mukohyama},\ and\ \citenamefont
  {Sotiriou}}]{Gumrukcuoglu:2016jbh}%
  \BibitemOpen
  \bibfield  {author} {\bibinfo {author} {\bibfnamefont {A.~E.}\ \bibnamefont
  {Gumrukcuoglu}}, \bibinfo {author} {\bibfnamefont {S.}~\bibnamefont
  {Mukohyama}}, \ and\ \bibinfo {author} {\bibfnamefont {T.~P.}\ \bibnamefont
  {Sotiriou}},\ }\href {\doibase 10.1103/PhysRevD.94.064001} {\bibfield
  {journal} {\bibinfo  {journal} {Phys. Rev.}\ }\textbf {\bibinfo {volume}
  {D94}},\ \bibinfo {pages} {064001} (\bibinfo {year} {2016})},\ \Eprint
  {http://arxiv.org/abs/1606.00618} {arXiv:1606.00618 [hep-th]} \BibitemShut
  {NoStop}%
\bibitem [{\citenamefont {{Peebles}}(1982)}]{1982ApJ...263L...1P}%
  \BibitemOpen
  \bibfield  {author} {\bibinfo {author} {\bibfnamefont {P.~J.~E.}\
  \bibnamefont {{Peebles}}},\ }\href {\doibase 10.1086/183911} {\bibfield
  {journal} {\bibinfo  {journal} {The Astrophysical Journal Letter}\ }\textbf
  {\bibinfo {volume} {263}},\ \bibinfo {pages} {L1} (\bibinfo {year}
  {1982})}\BibitemShut {NoStop}%
\bibitem [{\citenamefont {Trodden}(2016)}]{Trodden:2016zcu}%
  \BibitemOpen
  \bibfield  {author} {\bibinfo {author} {\bibfnamefont {M.}~\bibnamefont
  {Trodden}},\ }\bibfield  {booktitle} {\emph {\bibinfo {booktitle}
  {{Proceedings, 11th International Workshop on the Dark Side of the Universe
  (DSU 2015): Kyoto, Kyoto, Japan, December 14-18, 2015}}},\ }\href {\doibase
  10.22323/1.268.0005} {\bibfield  {journal} {\bibinfo  {journal} {PoS}\
  }\textbf {\bibinfo {volume} {DSU2015}},\ \bibinfo {pages} {005} (\bibinfo
  {year} {2016})},\ \Eprint {http://arxiv.org/abs/1604.08899} {arXiv:1604.08899
  [astro-ph.CO]} \BibitemShut {NoStop}%
\end{thebibliography}%

 \end{document}